\documentclass[twocolumn]{aastex631}

\newcommand{\Trat}{$T_{e^-}/T_p$} 
\newcommand{\DL}{$\Delta$log(z)}

\newcommand\ama[1]{\textcolor{black}{ #1}}

\usepackage{amsmath}
\usepackage{graphicx}
\usepackage{txfonts}
\usepackage{xcolor}
\usepackage{multirow,}
\usepackage{colortbl}
\usepackage{hyperref}
\usepackage{soul}
\usepackage{comment}

\begin{document}

\title{Evidence for thermal X-ray emission from the synchrotron dominated shocks in Tycho's supernova remnant}

\correspondingauthor{Amaël Ellien}
\email{a.r.j.ellien@uva.nl}

\author[ 0000-0002-1038-3370 ]{Amaël Ellien}
\affiliation{Anton Pannekoek Institute for Astronomy \& GRAPPA, University of Amsterdam \\
Science Park 904 \\ 1098 XH Amsterdam, The Netherlands }
\author[ 0000-0001-5792-0690 ]{Emanuele Greco}
\affiliation{Anton Pannekoek Institute for Astronomy \& GRAPPA, University of Amsterdam \\
Science Park 904 \\ 1098 XH Amsterdam, The Netherlands }
\author[ 0000-0002-4708-4219 ]{Jacco Vink}
\affiliation{Anton Pannekoek Institute for Astronomy \& GRAPPA, University of Amsterdam \\
Science Park 904 \\ 1098 XH Amsterdam, The Netherlands }
\affiliation{SRON Netherlands Institute for Space Research, Niels Bohrweg 4, 2333 CA Leiden, The Netherlands}

\begin{abstract}

Young supernova remnants (SNRs) shocks are believed to be the main sites of galactic cosmic rays production, showing X-ray synchrotron dominated spectra in the vicinity of their shock. While a faint thermal signature left by the shocked interstellar medium (ISM) should also be found in the spectra, proofs for such an emission in Tycho's SNR have been lacking. We perform an extended statistical analysis of the X-ray spectra of five regions behind the blast wave of Tycho's SNR using \textit{Chandra} archival data. We use Bayesian inference to perform extended parameter space exploration and sample the posterior distributions of a variety of models of interest. According to Bayes factors, spectra of all five regions of analysis are best described by composite three-component models taking into account non-thermal emission, ejecta emission and shocked ISM emission. The shocked ISM stands out the most in the Northern limb of the SNR.  We find for the shocked ISM a mean electron temperature $kT_{\rm{e}}=0.96^{+1.33}_{-0.51}$~keV for all regions and a mean ionization timescale $n_{\rm e}t=2.55^{+0.5}_{-1.22}\times10^9$~cm$^{-3}$s resulting in a mean ambient density $n_{\rm e}=0.32^{+0.23}_{-0.15}$~cm$^{-3}$ around the remnant. We performed an extended analysis of the Northern limb, and show that the measured synchrotron cutoff energy is not well constrained in the presence of a shocked ISM component. Such results cannot currently be further investigated by analysing emission lines in the 0.5-1 keV range, because of the low \textit{Chandra} spectral resolution in this band. We show with simulated spectra that X-IFU future performances will be crucial to address this point.

\end{abstract}

\keywords{Supernova remnants (1667), Interstellar medium (847), Bayesian statistics (1900), Posterior distribution (1926), Spectroscopy (1558)}


\section{Introduction}

\def\kms{{km\,s$^{-1}$}}

Observational results obtained over
the last two decades have greatly expanded our knowledge
of cosmic-ray acceleration by supernova remnants
\citep[SNRs, see][for reviews]{Reynolds2008,Helder2012,Vink2020}.
Apart from gamma-ray observations of SNRs, an important source of information on the cosmic-ray acceleration of properties of SNR shocks has been the detection of X-ray synchrotron emission from near the SNR shocks, which are caused by electrons with energies $\gtrsim 10$~TeV.

The evidence for synchrotron emission consist of the
nearly featureless
X-ray spectra from regions close to the shock fronts,
first established for SN\,1006 \citep{koyama95}, but now firmly
established for nearly all SNRs young than 1000--3000~yr \citep{Helder2012}.
In addition, for several young SNRs X-ray emission up to $\sim 100$~keV
has been established, in particular for Cas A and Tycho's SNR \citep{The1996,Allen1997, Favata1997, Vink2008, Grefenstette2015}.

Apart from nonthermal X-ray emission, young SNRs are in general
also emitting thermal emission from the hot, shock-heated plasma.
The thermal emission provides useful diagnostics about the electron $n_{\rm e}$ or ion $n_{\rm H}$ density,
the electron
temperature $kT_{\rm e}$, and how the ionisation is out of equilibrium,
characterised by the so-called ionisation age $n_{\rm e}t$, which itself
provides a measure of how long the plasma has been hot.

The combination of thermal and nonthermal X-ray emission provides us with information on both the
post-shock plasma properties as well as about the highest energy accelerated electrons. This is important, as SNR shocks
are collisionless shocks, in which the thermodynamic properties in the plasma at the shock are not established by
particle-particle collisions, like in the Earth atmosphere, but due to collective interactions, such as fluctuating electric and magnetic fields.
    One likely outcome of collisionless shocks is that the electron temperature may be much lower than the proton/ion temperature $kT_{\rm p}$. They may even have a ratio as low as  the ratio of the particle masses for Mach numbers $\gtrsim 40$, i.e. $kT_{\rm e}/kT_{\rm p}=m_{\rm e}/m_{\rm p}$ \citep{Ghavamian2013, Vink2015}.
Measuring the ion temperature is difficult, but can be done in X-rays under certain circumstances (e.g. \citealt{broersen13,mob19}), or using shocks moving through partially neutral gas using
Balmer-line diagnostics \citep[e.g.][for a review]{heng10}.

Ideally one would like to establish both the X-ray synchrotron and thermal X-ray emission properties from regions near the shock front in SNRs.
However, in practice this is hampered by the fact that many of the more prominent X-ray synchrotron filaments have spectra that seem almost devoid
of thermal X-ray emission, which is true for virtually all young SNRs, but an extreme example is RX~J1713-3946, whose X-ray emission is totally dominated by
synchrotron emission with hardly any thermal X-ray emission at all \citep{katsuda17}.

The question is now why there appears to be an anti-correlation between thermal and nonthermal X-ray emission?
On the one hand, one could reason that the X-ray synchrotron emitting plasma is in general associated with low density plasma,
suppressing thermal emission as this emission scales with $n_{\rm H}^2$. In some of the extreme cases like RCW 86 \citep[][]{vink06}, 
RX~J1713-3946 \citep{ellison12}, and G266.2--1.2 \citep["Vela Jr", e.g.][]{allen15} it could be that
the low density also resulted in a high shock velocity even at a SNR age of 1500--3000~yr, a prerequisite for X-ray synchrotron emission,
which requires shock velocities $V_{\rm s}\gtrsim 3000$~\kms\ \citep{aharonian99,Zirakashvili2007}.
However, this explanation does not fit well the case for young bright SNRs like Cas A, Kepler's SNR, and Tycho's SNR, as these are
known to evolve in relatively high density ambient media, and they are young enough to have high shock velocities despite these
high densities. 

Another explanation is that the electron temperature in the post-shock plasma of these young SNRs is very low, which could be either
due an extremely low ratio $kT_{\rm e}/kT_{\rm p}$, or even due to extremely non-linear DSA, in which most shock energy is diverted
to CR acceleration rather than the thermal energy of the post-shock plasma (\citealt{Giuffrida2022}). For example, \citet{drury09} suggested that plasma temperatures
may be as low as $\sim 6$ times the upstream temperature, suggesting a temperature that could be as low as 30,000--60000~K (2.6--5.2~eV). However, taking into account all thermodynamic relations does not allow for such low plasma temperatures,
and for a CR efficiency of $w=25$\% a reduction in plasma temperature of $\sim 30$\% is more reasonable \citep{vink10a}.

Here we report on a study of these regions in Tycho's SNR (the remnant of the Type Ia supernova of 1572) whose X-ray  emission seem to be completely dominated  by synchrotron radiation \citep[e.g.][]{Warren2005,Cassam2007}.
We setup this study with the goal of detecting (hints of) thermal emission, and to determine what this implies
for the local conditions of the plasma {and for the ambient medium of the remnant. The study itself was carried out using archival Chandra  X-ray data.
In addition, we want to reassess whether the putative thermal emission can be  measured with the future high spectral resolution X-ray spectrometer X-IFU on board Athena
\citep{Barret2018}, to not only measure local densities and electron temperatures, but also the ion temperature through Doppler broadening.
The latter can be used to also put limits on the electron to ion temperature ratio, as well as on the temperature reduction that might be expected if shock acceleration is very efficient.

In Section~\ref{sec:observation_and_data_reduction}, we describe the data set we use for the study and the reduction procedure we followed to obtain our spectra. In Section~\ref{sec:statistical_analysis}, we give some context on Bayesian analysis contrasted with previous studies on SNR's shocks, and describe the different models of interest we use for the analysis. In Section~\ref{sec:simulations}, we use maximum posterior parameters of the best Bayesian selected model to produce simulated spectra of X-IFU. We then test the detectability of Doppler broadening on emission lines in these simulations. In Section~\ref{sec:conclusions} we present our conclusions.

   \begin{figure*}
       \plotone{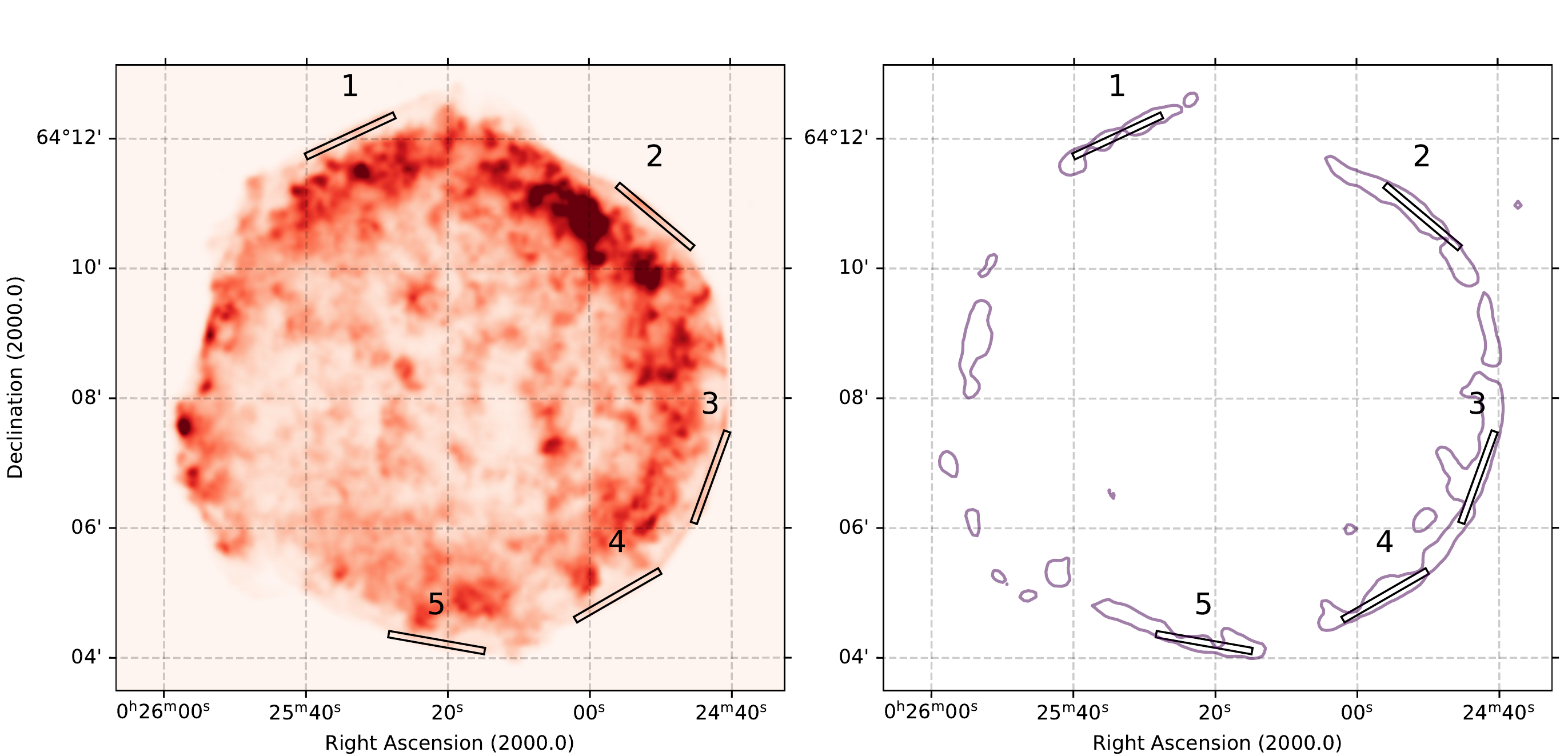}
       \caption{Left: broadband \textit{Chandra} image of Tycho's SNR. Right: schematic view of Tycho's SNR. The five black boxes are the regions over which our shock spectra were extracted and analyzed (see Section~\ref{sec:observation_and_data_reduction}). The contours are drawn from the contrast image computed from the normalized 1.7-1.95 keV and 4.0-6.0 keV images. Note that the contrast image has been smoothed with a $1\sigma$ Gaussian kernel before drawing the contours.}
       \label{fig:tycho_flux_image}
   \end{figure*}

   \begin{figure}
       \plotone{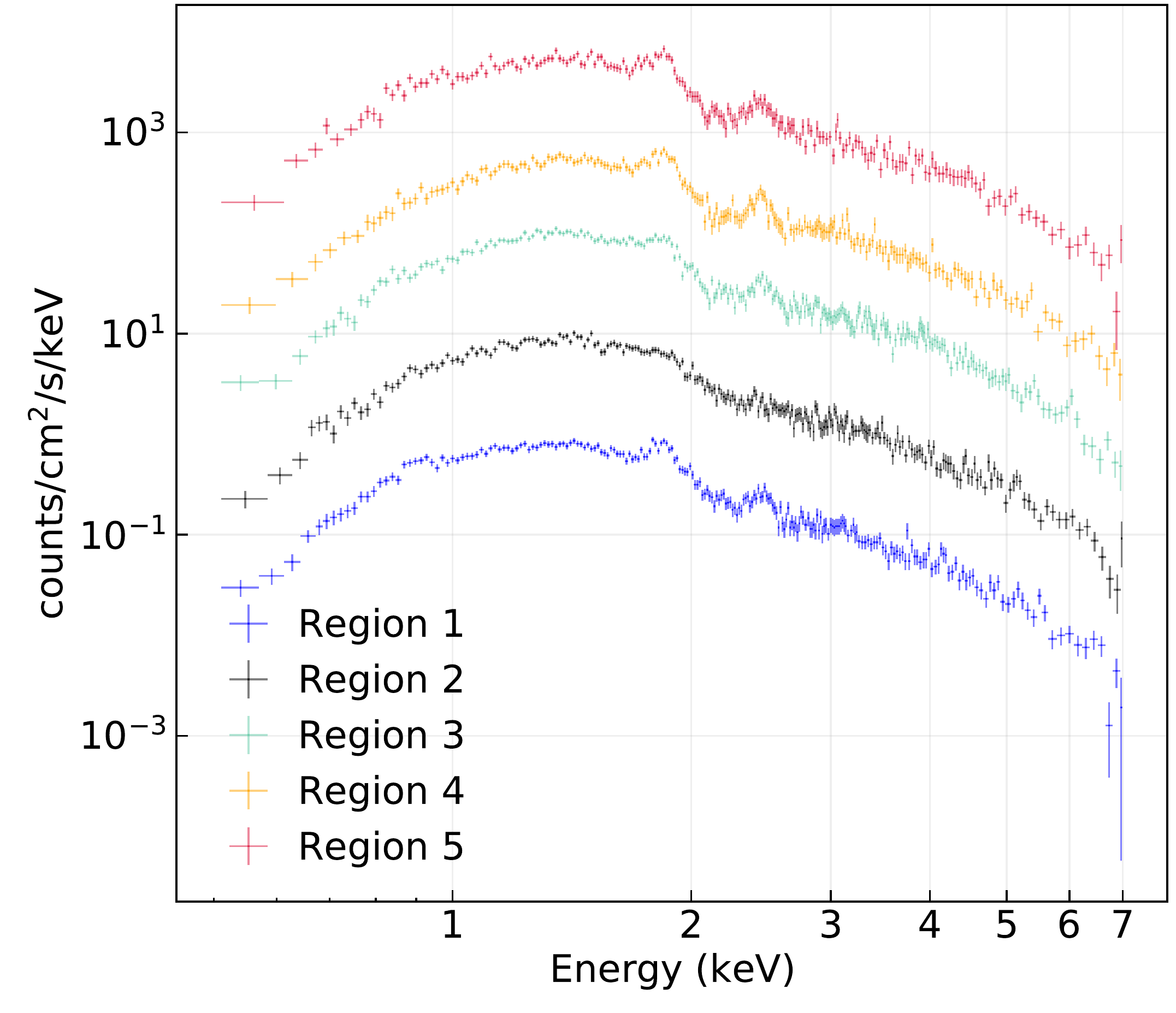}
       \caption{Spectra in the 0.5-7.0~keV band of the five regions. Region~1 has native normalization, but every other spectra are renormalized by a factor 10 for display purposes.}
       \label{fig:tycho_flux_spectra}
   \end{figure}

\section{Observations and data reduction}
\label{sec:observation_and_data_reduction}


We use archival \textit{Chandra} data for our analysis and pick the longest single observation available for the Tycho's SNR (ObsId 10095, PI: J. P. Hughes). This observations was made in April 2009 with the Advanced CCD Imaging Spectrometer (ACIS) using the four ACIS-I front illuminated chips. The field of view (FoV) allows to cover the full remnant, with a total integration time of 173.37 ks. We use the Chandra Interactive Analysis of Observations (\href{https://cxc.cfa.harvard.edu/ciao/}{\texttt{CIAO}}, version 4.12) and the Calibration Database (\texttt{CALDB}, version 4.9.3) to reprocess the data following the standard \textit{Chandra} procedure through the task \texttt{chandra\_repro}. We also use the \texttt{blanksky} command line, which takes \texttt{CALDB} blank sky exposures matching our current data set to produce the background file for our study. This gives a larger area to extract the background spectrum for our spectral analysis than a local background estimation, as Tycho's SNR occupies most of the FoV. The \texttt{fluximage} command line is used to create a broadband flux image as well as 1.7-1.95 keV and 4.0-6.0 keV images. The 1.7-1.95 keV band covers the Si lines, therefore tracking the bulk of the ejecta emission. The 4.0-6.0 keV band is on the other hand a continuum band without any dominant emission line, typically used to track the synchrotron emission. 

To proceed to the study of Tycho's blast wave, we select spatial regions very close to the shock front. These regions must be large enough to provide enough counts for the analysis, while still avoiding as much as possible ejecta emission. To do so we normalize to unity the 1.7-1.95 keV and 4.0-6.0 keV images, and subtract them from each other to create a contrast image. This allows us to exacerbate the spatial regions strongly continuum dominated with minimum contamination from the ejecta. We use this information as well as the broad band image to draw by hand five thin boxes of size $90\arcsec\times6\arcsec$ over spatial areas of interest. The broadband images as well as the regions on top of contrast contours are displayed in Figure~\ref{fig:tycho_flux_image}. Despite this procedure, we still expect a few amount of ejecta to be covered by our regions and slightly contaminating our data. Since the size of the regions considered is significantly higher than the nominal \textit{Chandra}/High Resolution Mirror Assembly PSF (HRMA), of the order of 0.''5 \footnote{Table 4.1 in \href{https://cxc.harvard.edu/proposer/POG/html/chap4.html}{Chandra OG website}} we can ignore any PSF effect in our analysis.

We extract the spectra in each region using the \texttt{specextract} command line and the blank sky background file, with a binning of at least 30 counts per bin. The spectra are displayed in Figure~\ref{fig:tycho_flux_spectra} and their properties in Table~\ref{tab:regions_props}. 
Note that the spectra of all regions except region 2 display visible signs of Si emission lines, likely due to ejecta emission. Given the purpose of this paper, Region~2 is chosen to be our main analysis region as it provides the cleanest case for a synchrotron dominated X-ray spectrum, while the others are kept as supplementary information sources.


\begin{tiny}
    \begin{table*}
    \centering
    \begin{tabular}{c c c c c c c}
    \hline\hline
    Name & Right Ascension & Declination & Length & Width & Angle & Number of counts \\
     & [J2000] & [J2000] & [$\arcsec$] & [$\arcsec$] & [°] & [ counts ]\\
    \hline
     Region~1 & 0:25:33.8304 & +64:12:02.980 & 90 & 6 & 25 & 14344  \\ 
 Region~2 & 0:24:50.6890 & +64:10:48.254 & 90 & 6 & 320 & 24338  \\ 
 Region~3 & 0:24:42.9364 & +64:06:47.149 & 90 & 6 & 70 & 21082  \\
 Region~4 & 0:24:56.0738 & +64:04:58.050 & 90 & 6 & 30 & 20243  \\ 
 Region~5 & 0:25:21.5883 & +64:04:14.491 & 90 & 6 & 170 & 14494  \\ 
\hline
\\
    \end{tabular}
    \caption{Properties of analysis regions.}
    \label{tab:regions_props}
    \end{table*}
    \end{tiny}

\section{Statistical analysis}
\label{sec:statistical_analysis}

\subsection{Context}
\label{subsec:context}



The successful (or not) detection of a thermal component in SNR's shock wave spectra is diagnosed by model comparison. A non-thermal model giving a statistically better fit than a thermal one would strongly point toward a non-thermal origin for the spectrum. This is the case for numbers of SNRs, where the featureless shock spectra are better represented by power-law distributions or models that take into account the curvature
expected for hard X-ray synchrotron radiation \citep{Bamba2005, Helder2008}. In the same manner, a composite model with two components (thermal plus non-thermal), providing a significantly better fit than a single non-thermal component model, would imply that a thermal feature is detected in addition to the dominating non-thermal component.

However some caveats must be underlined when using such procedures. 
The statistic mostly used to estimate the quality of a fit has been the $\chi^2$-statistic, a Gaussian approximation of X-ray emission processes following Poisson distributions. This approximation is applicable in a very high count number regime, which is usually not the case in X-ray astronomy. Studies of SNR's shock has not been short of using this approximation in a variety of works. Another statistic is the \textit{C}-statistic \citep{Cash1979}, also called \textit{C-stat}. The \textit{C-stat} is based on a Poisson likelihood, and therefore is more appropriate for X-ray spectral analysis. 

Aside from the statistic used to diagnose fit quality, the way comparisons has been done to discriminate between different models is also problematic. At best no statistical comparison is done between fits using two different models and conclusions are drawn independently in both cases. Often the difference of reduced $\chi^2$-statistic or \textit{C-stat} is used as metric to qualify which model is the best one to represent the spectrum. It has been shown than such statistical tests were only adequate for embedded models with parameter values far from boundaries \citep{Protassov2002}. This is not the case when comparing thermal, non-thermal models or combinations of both. Using another metric for model comparison is therefore an important matter to correctly assess the detection or not of a thermal component.

Last but not least, fitting algorithms are local optimization algorithms. When fed starting parameter values, these algorithms will iterate over the parameter space trying to minimize (maximize) the statistic (likelihood). After iterating, they give the single set of parameter value corresponding to the likelihood maximum over the given parameter bounds. Such algorithms perform well in simple cases when the likelihood distribution of the model over the data is monomodal. However, they cannot account for complex parameter spaces with multiple likelihood peaks, for instance. Therefore, is it possible to ensure that the parameter space of our model over the data has been correctly explored? One cannot just explore manually highly dimensional parameter spaces, due to the curse of dimensionality and computation time. These and other similar issues are limits to our comprehension of the underlying physics and can be addressed by using more robust statistical tool.

\subsection{Bayesian X-ray Analysis}
\label{subsec:bayesian_xray_analysis}

The problems highlighted in the previous section are well known nowadays and are not limited to the study of SNR's shocks. Dubbed as \textit{frequentist} inference, this procedure has been contrasted many times with another class of methods denoted as \textit{Bayesian} inference \citep{vanDyk2001}. In the Bayesian framework, the parameter space is still explored using the likelihood distribution. However rather than trying to find likelihood maxima, this approach consists in evaluating the posterior probability distribution of the model over the data. It is done by deforming the parameter space with priors and identifying regions of it enclosing most of the information, e.g. regions of the parameter space in which most of the likelihood resides. It is insensible to whether the likelihood distribution is mono or multimodal, and allows to estimate at the same time parameter values and their uncertainties.

In addition to parameter estimation, Bayesian inference allows model comparison. Instead of comparing goodness-of-fit statistics as in frequentist model comparison, this approach integrates the likelihood over the parameter space to compute the marginalized likelihood, also called Bayesian evidence. The ratio of evidences of two different models is the Bayes factor \citep{Kass1995, Trotta2008}, which is an efficient metric to estimate which model better represents the data. The most common scale used to estimate the significance of a Bayes factor is Jeffrey's scale \citep{Jeffreys1961}, that we describe in the Section~\ref{subsec:metrics}.

In comparison to frequentist approaches, Bayesian methods are even more prone to computing time issues as the likelihood cannot be estimated on a grid over the whole parameter space. An approximation must be computed, a procedure denoted as \textit{sampling}. To do so efficient methods have been proposed, such as nested sampling \citep{Skilling2004, Buchner2021a}. We will not go into details about this method, and only highlight the fact that it allows to produce posterior samples containing most of the information of the true (multimodal or not) likelihood distribution, and to compute the Bayesian evidence with an error estimate. An efficient and robust implementation of nested samplings has been done with the \texttt{Ultranest} algorithm \citep{Feroz2009}. More recently, \texttt{Ultranest} has been linked to the classic X-rays analysis package \texttt{XSPEC} \citep{Arnaud1996} through the user friendly Python package \texttt{BXA} \citep{Buchner2014}. A complete X-ray spectral analysis in a Bayesian framework is therefore made possible, using the standard \texttt{XSPEC} models. The versions used in this work were \texttt{XSPEC} v12.11.1 and \texttt{BXA} v4.0.2.    

\subsection{Metrics}
\label{subsec:metrics}


The Bayesian evidence computed by \texttt{BXA} is an approximation of the true, continuous Bayesian evidence. Therefore it comes with a budget error for each model, which is around $\sim0.5$ for log(z) in the case of our spectra and models. This error propagates to Bayes factor and results in an error ranging from 0 to $\sim1$, which is of the same order as the threshold \DL~$\sim2$ and could in some cases strongly impact the interpretation. These uncertainties must be kept in mind when comparing and assessing the models, and we usually considered greater threshold values for \DL~ for the Bayes factor to significantly discriminate between two models.

Additionally, while the Jeffrey's scale is a good first order estimation of the significance of Bayes factors, one must remember that it is an arbitrary set of values for such purpose. In order to truly estimate the significance of a Bayes factor, one would need to compute its distribution over each specific dataset using Monte Carlo realizations of mock spectra \citep{Keeley2022}. However this is far outsides the bounds of what is realistically possible in our case, as it took weeks to sample the posteriors for some of our models, even when parallelized on several dozens of computing cores. Regarding these technical constraints and for simplicity, we still resorted to Jeffrey's scale to make sense of our Bayes factors and evaluate the associated models.

In this work we used a series of complementary metrics to i) assess a model's capacity to reproduce accurately the observed spectra and ii) perform model selection. For case i) we used the \textit{C-stat} metric (and its reduced version denoted $C_r$), as it has been traditionally used for such purpose in the literature. In our case, the $C_r$ acted as a safeguard metric which allowed us to make connections between our newly used Bayesian inference metrics and more frequently used frequentist properties. For case ii) we used as main metric for each pair of model A and B the Bayes factor, denoted \DL~such as \DL~= log(z)$_A$ - log(z)$_B$. Both log(z)$_A$ and log(z)$_B$ are outputs of \texttt{BXA}. To interpret these values, we used the Jeffreys's scale \citep[][]{Jeffreys1961} which states that a $\mid $\DL$ \mid $~$\ge 2$ is 'decisive' (e.g. that the model with the highest evidence is decisively better at representing the data). On the other hand, a Bayes factor $\mid $\DL$ \mid $~$< 2$ signifies that the two models cannot be significantly distinguished.

In order to double check our results and  add robustness to our model selection, we also computed three additional metrics based on \textit{C-stat}, similarly to what was done in \citet{Buchner2014}. The first one is the Akaike Information Criterion \citep[AIC;][]{Akaike1974}. The AIC is given by $AIC=\mathrm{\textit{C-stat}}-2m$ with $m$ the number of model parameters. Such a metric measures the loss of information when using a model. The model with the lowest $AIC$ is therefore the best model to describe the data. The second metric is the Bayesian Information criterion \citep[BIC;][]{Schwarz1978}, given by $BIC=\mathrm{\textit{C-stat}} - m\ln(n)$, with $n$ the number of degrees of freedom. The BIC is an approximation of Bayesian model selection and assumes a very peaked likelihood maximum, making the priors negligible. Therefore it only uses the likelihood maximum to assess which model better reproduces the data. The model with the lowest BIC should be preferred over the others. The last metric we used is the difference in \textit{C-stat} of two models, denoted $\Delta C=\mathrm{\textit{C-stat}}_A - \mathrm{\textit{C-stat}}_B$. We would like to stress once more that the use of these three metrics makes sense only as additional supports for the Bayes factor, as they are all based on strong (and not necessarily true) assumptions about the likelihood distribution and the models.

\subsection{Single-component models}
\label{sec:single_component_models}
\begin{figure*}
\centering
\plotone{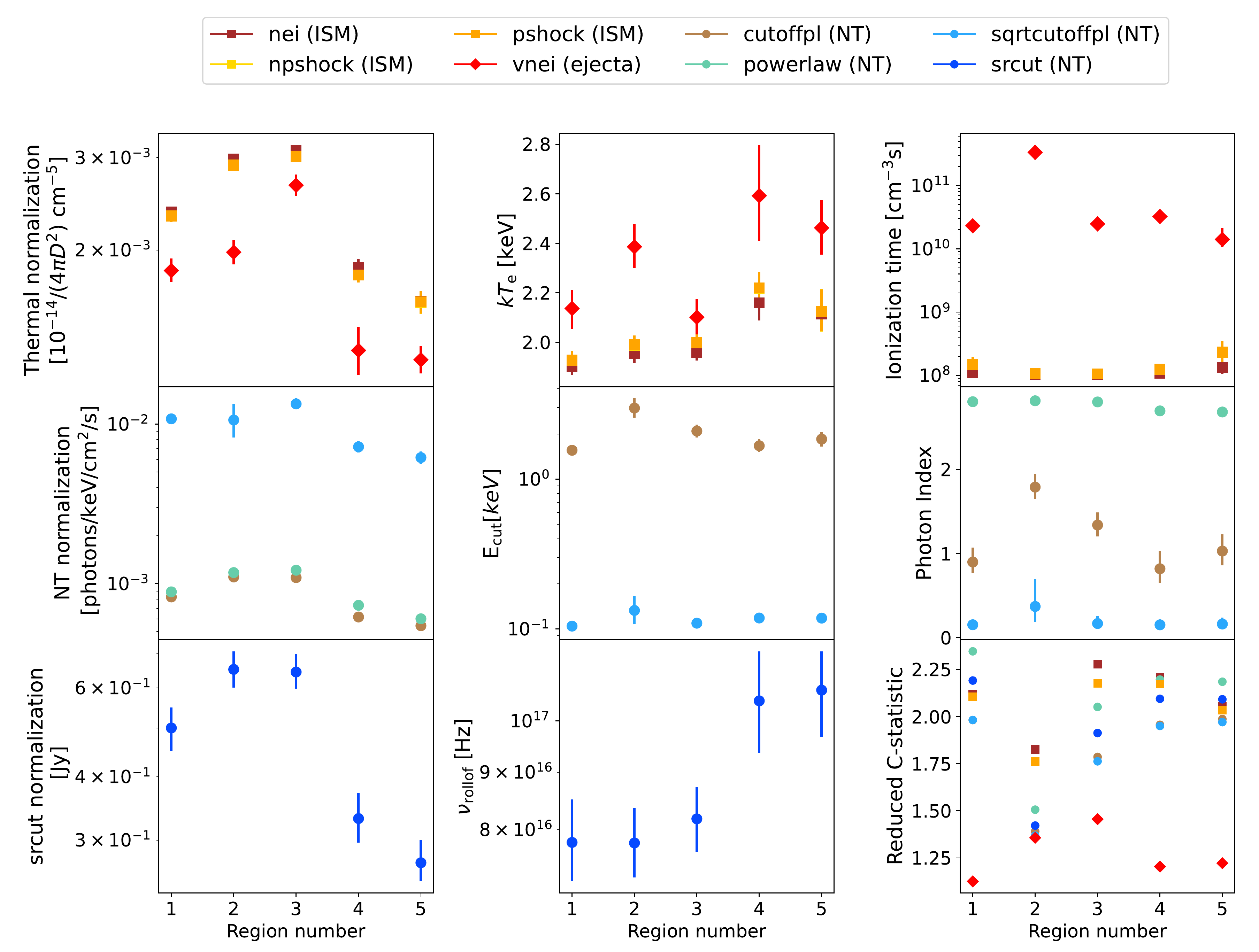}
\caption{Best Bayesian parameter values with 1-$\sigma$-equivalent quantile error bars for single-component models in each of the five regions. Only the main parameters and the \textit{C-stat} are shown. The physical component of each model is indicated in the legend (NT stands for non-thermal). Some error bars are smaller than the corresponding data points and are therefore not visible. The \texttt{npshock} data points are not visible because strongly overlapping with the \texttt{pshock} data points.}
\label{fig:plot_single_component_values_3by2}
\end{figure*}

\begin{figure}
\centering
\plotone{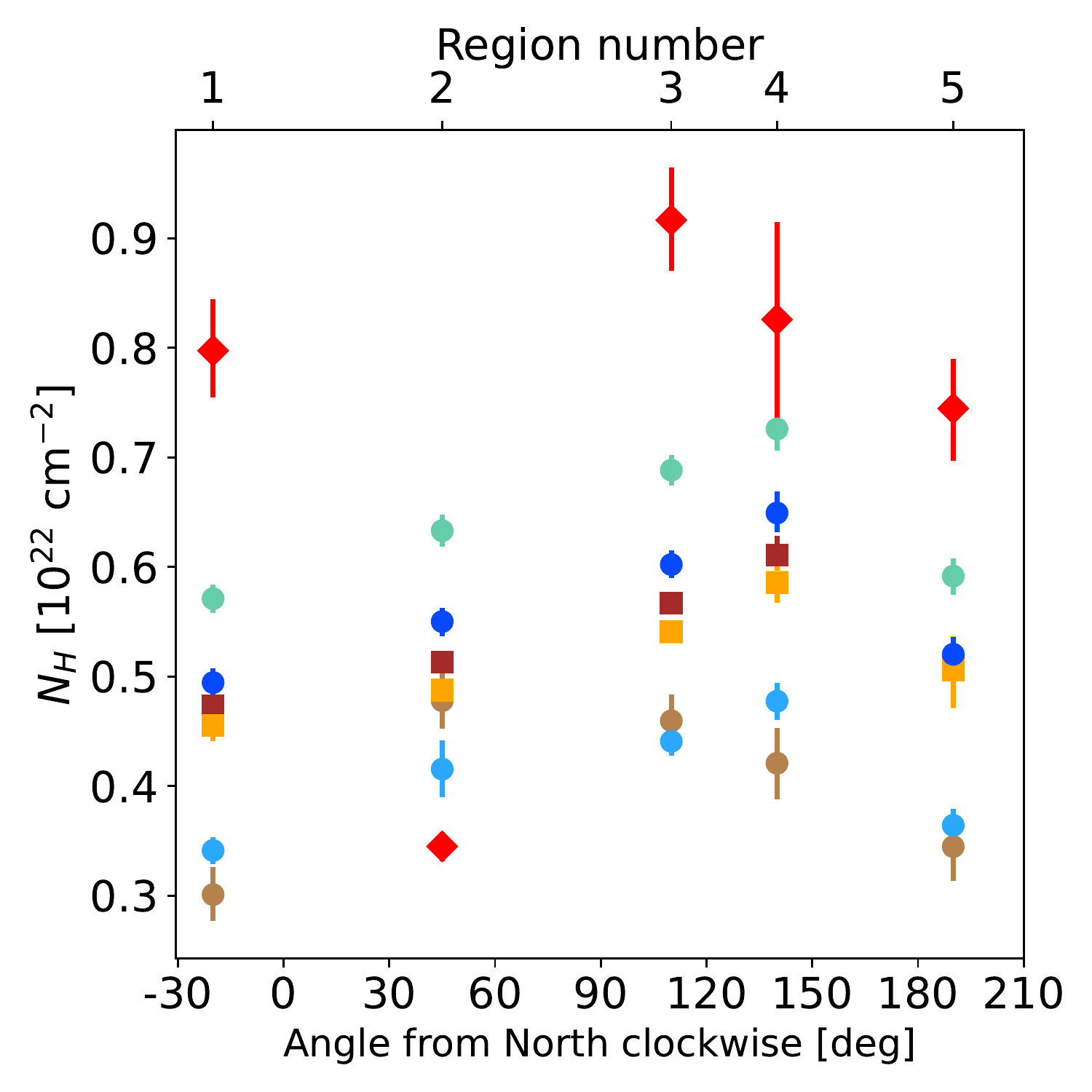}
\caption{Best Bayesian $N_H$ values with 1-$\sigma$-equivalent quantile error bars for single-component models in each of the five regions. The color code is the same as Figure~\ref{fig:plot_single_component_values_3by2}. Some error bars are smaller than the corresponding data points and are therefore not visible.}
\label{fig:plot_nh_single_comp}
\end{figure}


\begin{table*}
\caption{Bayes factors for all single component models models in each of the five regions.}
\label{tab:regions_logz_single_comp_models}
\centering
\resizebox{\textwidth}{!}{
\begin{tabular}{l l c c c c c c}
\hline\hline\\
& & \multicolumn{6}{c}{log(z) - log(z)$_\mathrm{max}$} \\\\\cline{3 - 8}\\
& Model & Region 1 & Region 2 & Region 3 & Region 4 & Region 5 & Average \\\\
\hline\\
\multirow{4}{*}{Single non-thermal}
& \texttt{TBabs(powerlaw)} & -254.45  & -29.00  & -116.39  & -205.27  & -200.08  & -159.84 \\
& \texttt{TBabs(srcut)} & -221.15  & -11.94  & -86.94  & -184.05  & -181.02  & -135.82 \\
& \texttt{TBabs(cutoffpl)} & -176.72  & -6.82  & -60.70  & -154.55  & -159.00  & -110.36 \\
& \texttt{TBabs(sqrtcutoffpl)} & -176.93  & \cellcolor{green}0.00  & -54.78  & -152.83  & -154.82  & -106.67 \\
\\
\hline
\\
\multirow{3}{*}{Single thermal ISM}
& \texttt{TBabs(nei)} & -208.98  & -105.72  & -172.72  & -212.21  & -174.81  & -173.69 \\
& \texttt{TBabs(pshock)} & -203.23  & -90.04  & -149.00  & -202.59  & -168.19  & -161.41 \\
& \texttt{TBabs(npshock)} & -203.64  & -89.63  & -148.39  & -202.68  & -168.62  & -161.39 \\
\\
\hline
\\
\multirow{1}{*}{Single thermal ejecta}
& \texttt{TBabs(vnei)} & \cellcolor{green}0.00  & -6.00  & \cellcolor{green}0.00  & \cellcolor{green}0.00  & \cellcolor{green}0.00  & \cellcolor{green}0.00 \\
\\
\hline
\end{tabular}}

\tablenotetext{}{Notes. For each region (column) the Bayesian evidence of every model is normalized to the best one in that region (column). In the last right column is shown the Bayesian evidence for each model averaged over all regions, normalized by the best average evidence. The best model for each region has a value of zero and is highlighted in green. Models with very low $\Delta \mathrm{log(z)}$ are the models providing the worst representations of the data (which is in average the \texttt{TBabs(nei)} model here).}
\end{table*}

The first part of our spectral analysis was concentrated on single component models in order to check the consistency between the results we get in this paper with the \texttt{BXA} package with those reported in previous work. In the following, all models were coupled to a Tuebingen-Boulder interstellar medium (ISM) absorption model \citep[][defined in \texttt{XSPEC} as \texttt{TBabs}]{Wilms2000}. Hydrogen column density values of $\sim0.7\times10^{22}$~cm$^{-2}$ were reported by \citet{Cassam2007}, and to keep the analysis more general we set the boundaries between 0 and $2\times10^{22}$~cm$^{-2}$ with linear prior for all models.}

\subsubsection{Synchrotron emission}

The non-thermal radiation is emitted by relativistic electrons having a power-law distribution with an high energy exponential term $\exp\left[-\left(E/E_{\mathrm{cut}}\right)^{\alpha}\right]$. Such electron spectra result in synchrotron emission following a power law with an exponential cutoff $\exp\left[-\left(E/E_{\mathrm{cut}}\right)^{\beta}\right]$. Depending on whether the electron distribution is radiative loss-limited \citep[$\alpha = 2$, giving the so called power-law with exponential cutoff electron spectrum;][]{Zirakashvili2007} or age-limited \citep[$\alpha=1$;][]{Reynolds1999} directly influences the parameter $\beta$ and the steepness of the cutoff. For example, \citet{Zirakashvili2007} performed an analytical treatment of the shock-accelerated electron spectrum and showed that $\beta = 1/2$ in the case of Bohm diffusion and $\beta=1/3$ in the idealized case of energy-independent diffusion. The most-used \texttt{XSPEC} model for synchrotron spectra, \texttt{srcut} \citep{Reynolds1999}, assumes an electron spectrum with an exponential cutoff and uses approximations to derive the associated photon spectrum. This also results in $\beta = 1/2$, although the starting electron distribution hypothesis is different and incompatible with results from \citet{Zirakashvili2007}. Nevertheless the \texttt{srcut} has been used to derive the relevant cutoff photon energy of synchrotron spectra \citep{Bamba2005, Lopez2015}.

We considered a \texttt{srcut} model as well as a \texttt{cutoffpl} model, namely a power law with an exponential cutoff given by $\exp(-E/E_{\mathrm{cut}})$. The latter results in a steeper spectrum than the regular power law at high energies. We also included a custom modified cutoff model, which we named \texttt{sqrtcutoffpl}. This is the exact same as the \texttt{cutoffpl} model, but the exponential cutoff is given by $\exp(-\sqrt{E/E_{\mathrm{cut}}})$. The \texttt{sqrtcutoffpl} model is modeling solutions from \citet{Zirakashvili2007, Zirakashvili2010}, and gives a a more gradual break at high energy. These three models, in addition to the power-law, allow us to probe extensively the effect of various cutoffs on the description of the spectrum. We refer to these as 'non-thermal' models or components for the rest of this work.

The parameters of non-thermal models were allowed to vary over the whole parameter space set by \texttt{XSPEC} boundaries, with the exception of the cutoff energy that was set in the range $\left[10^{-1}, 10 \right]$~keV. Log-uniform priors were given to all normalization parameters, as well as to the break frequency $\nu_{\mathrm{rolloff}}$ for the \texttt{TBabs(srcut)} model and to the cutoff energy for the \texttt{cutoffpl} and \texttt{sqrtcutoffpl} models. For every other parameter, uniform priors in linear space were assumed.

The resulting photon indexes and normalizations for the powerlaw and its variants for all regions are displayed in Figure~\ref{fig:plot_single_component_values_3by2}. For the simple power-law model, we find $\Gamma \sim 2.7 - 2.9$ which is similar to what was found in the vicinity of the shock by \citet{Cassam2007}. Note that the photon index value strongly varies depending on the model, with $\Gamma \sim 0.7 - 1.9$ for the \texttt{TBabs(cutoffpl)} model and $\Gamma \sim 0.15 - 0.4$ for the \texttt{TBabs(sqrtcutoffpl)} model. The same can be said about the normalization of these models with \texttt{TBabs(cutoffpl)} values higher than the other two by one order of magnitude. The \texttt{srcut} break energy and its normalization are also displayed in two separate frames. The values are consistent with values found in the literature \citep{Bamba2005}.

The $N_H$ values of all non-thermal models are shown in Figure~\ref{fig:plot_nh_single_comp}. The values for the \texttt{TBabs(powerlaw}) and \texttt{TBabs(srcut)} are once again consistent with values from \citet{Bamba2005} and \citet{Cassam2007}. This is not the case for cutoff models showing lower $N_H$ values ($\sim0.3-0.5\times10^{22}$~cm$^{-2}$) than the powerlaw ($\sim0.6-0.75\times10^{22}$~cm$^{-2}$).

The Bayesian factors are displayed in Table~\ref{tab:regions_logz_single_comp_models}. The simple power-law gives the worst representation, as all other non-thermal models have better Bayes factors for all regions. This is consistent with the cutoff expected from synchrotron emission \citep{Zirakashvili2007}.

\subsubsection{Shocked interstellar medium emission}

In the literature, shocked ISM emission is usually described with non-equilibrium of ionization (\texttt{nei}) models \citep{Hwang2002, Bamba2005} with a few exceptions such as the cosmic ray modified model from \citet{Cassam2007}. In addition to the \texttt{nei} model, we also perform the analysis with plane-parallel shock models, both with single temperature (\texttt{pshock}) and with separate electron and ion temperatures (\texttt{npshock}),
see \citet{Borkowski2001}. These models take into account the gradient in ionisation age downstream of the shock, with an upper value ($n_{\rm e}t)_u$ and a lower value ($n_{\rm e}t)_l$, rather than considering the mean value of $n_{\rm e}t$ for the entire shock. The \texttt{npshock} model, in addition, also
takes into account the slow, collisional equilibration of the electron temperature due to  Coulomb interaction between the protons and the electrons, assuming $kT_{\rm e}<kT_{\rm p}$ at the shock. We refer to this family of models as 'ISM' for the rest of this work.

For the \texttt{npshock} model, we took shock velocity measurements from \citet{Williams2016} to compute the mean shock temperatures in our different regions and provide them as input parameters, while the electron temperature is left free. We assumed a distance of 2.3~kpc for Tycho's SNR to compute these values (see Table~\ref{tab:williams_temperatures} for shock velocity and mean temperature values). The parameters were allowed to vary over the whole parameter space set by \texttt{XSPEC} boundaries. Log-uniform priors were given to all normalization parameters, and to the ionization timescale. The abundance was frozen to solar abundances \citep{Anders1989} to replicate what was done in the literature. For every other parameter, uniform priors in linear space were assumed.

The resulting likelihood posterior distribution for each model is monomodal and Gaussian-like, so we do not display it here. The best fit parameter values are shown for all regions in Figure~\ref{fig:plot_single_component_values_3by2}. The best-fit parameter values for the \texttt{TBabs(pshock)} and \texttt{TBabs(npshock)} models are so close that some of the data points in the corresponding frame are not visible because they are overlapping. We found $kT_{\rm e}\sim2$~keV in all regions for \texttt{TBabs(nei)}, \texttt{TBabs(pshock)} and \texttt{TBabs(npshock)} models, which is similar to the values found by \citet{Hwang2002} and \citet{Warren2005} for the electron temperature. The ionization timescale of all three models show values around $10^8$~cm$^{-3}$s. The $N_H$ values are also displayed in Figure~\ref{fig:plot_nh_single_comp}, and are also very similar for all three models with $N_H\sim0.6\times10^{22}$~cm$^{-2}$.

The Bayes factors of the ISM models are displayed in Table~\ref{tab:regions_logz_single_comp_models}. The \texttt{TBabs(pshock)} and \texttt{TBabs(npshock)} models have similar Bayes factors, which are significantly better than the standard \texttt{TBabs(nei)} model.


\begin{table}

    \caption{Shocked region properties from \citet{Williams2016}. }
    \centering
    \begin{tabular}{c c c c}
        \hline\hline
        Region number & Region number & $V_S$ & $kT_{\rm e}$ \\
        $\left(\mathrm{This~work}\right)$ & (Williams 2016) & [km.s$^{-1}$] & [keV] \\
        \hline
        1 & 1 & 2891 & 9.77 \\
        2 & 16 & 3328 & 12.95 \\
        3 & 12 & 3644 & 15.53 \\
        4 & 11 & 3666 & 15.72 \\
        5 & 9 & 3633 & 15.44 \\
        \hline
        \\
    \end{tabular}
    
    \label{tab:williams_temperatures}
    \tablenotetext{}{Notes. First column displays the number of the analyzed regions in this work, second column the number of the closest corresponding regions from \citet{Williams2016}, third column the shock velocity computed from their expansion rates, and fourth column the associated mean shock temperature \citep[see Equation~4.12 in][]{Vink2020}. Note that these values have been derived assuming a distance of 2.3~kpc for Tycho's SNR.}
    
\end{table}

\subsubsection{Shocked ejecta emission}

In order to correctly assess the nature of the spectra, one must look for the presence of ejecta emission. This is a delicate task, as there is not much thermal features in the spectra, which makes it hard to constrain ejecta properties. The ejecta component was modeled using a \texttt{vnei} model. The plasma temperature was allowed to vary between 1 and 5 keV and the ionization timescale  over the whole parameter space. The H, He, C, N and Ni abundances were frozen, and the other elements free to vary between $10^{-2}$ and $10^{1}$ with log priors.

The resulting posterior distributions are monomodal for most parameters, although for all models a few abundances are poorly constrained. Most notably the Ne, Ca and Ar abundances display either flat parameter spaces, or their distribution is cut by lower parameter value boundaries. This is enhanced in Region~2, where no element is correctly constrained barring the Si abundance. All the best-fit abundances values are unrealistically low (e.g. $\leq 0.4$, with the highest values being $\sim0.4$ for the Si and the S), and cannot be meaningfully interpreted as representing  ejecta components.
The best-fit values for the main parameters ($kT_{\rm e}$, $n_et$ and the normalization) are displayed in Figure~\ref{fig:plot_single_component_values_3by2}. All values are clearly different from the ISM models best-fit values, with notably higher electron temperatures ranging between 2.1 and 2.6~keV, ionization timescale one order of magnitude greater with values around $10^{10}$cm$^{-3}$s, and even as high as $3\times10^{11}$cm$^{-3}$s for Region~2.

The Bayes factor or the \texttt{TBabs(vnei)} model for all regions are shown in Table~\ref{tab:regions_logz_single_comp_models}. Please note that this model was the longest single component model to process due to the higher number of free parameters and the low amount of discriminating features in the spectra to fit these parameters.



\subsubsection{Comparison of single component models}
\label{subsub:comparison_single_component_models}

As comparison metric we used the Bayesian evidence log(z) computed by \texttt{BXA} (see Section~\ref{subsec:metrics} for details). For each region the highest evidence was used as a normalization. The resulting Bayes factors \DL~= log(z) - log(z)$_{\mathrm{max}}$ can be found in Table~\ref{tab:regions_logz_single_comp_models}, where they are shown for all regions. In the same table the log(z) for each model averaged over all regions is also shown, normalized by the best average. To interpret these values, we used the Jeffreys's scale \citep[][]{Jeffreys1961} which states that a $\mid $\DL$ \mid $~$\ge 2$ is 'decisive' (e.g. that the model with the highest evidence is decisively better at representing the data). The best model for each region spectrum has therefore a value \DL~equal to zero, and the lowest the Bayes factor is for a model, the less appropriate it is to describe the spectra. We highlight the highest evidence models in green in the table.
In the following paragraphs, we also use this threshold $\mid $\DL$ \mid $~$\sim 2$~to compare other inadequate models together, as there are interesting points to make about these too. We also computed the AIC, BIC and $\Delta C$ for all models (see Section~\ref{subsec:metrics} for a description of these metrics). In the same way as the Bayes factors, they were computed for each model and normalized by the best model. All values are shown respectively in Tables~\ref{tab:aic}, \ref{tab:bic} and \ref{tab:deltac} in appendix, with the best model highlighted in green.

For four out of the five regions, the thermal \texttt{vnei} model gives the best representation according to Bayes factors, demonstrating the importance of a refine modeling of ejecta. This is also visible in the \textit{C-stat} values, as shown in the lower right panel of Figure~\ref{fig:plot_single_component_values_3by2}.

The exception is Region~2, which is better represented by our custom cutoff power-law (\texttt{sqrtcutoffpl}). This supports our preliminary assertion about Region~2 being the least contaminated by ejecta and the best candidate region to study shocked ISM (see Section~\ref{sec:observation_and_data_reduction} and Figure~\ref{fig:tycho_flux_spectra}).

However, no single-component model is able to correctly capture the physical parameters of a complex multi-component plasma, as shown by the unrealistically low abundances given by the \texttt{vnei} model, or by the goodness-of-fit metrics. A combination of these different components is needed to get an acceptable representation of these spectra.

\subsection{Two components models}
\label{sec:bicomponent_models}

In this section we combine a non-thermal component alongside a thermal one, which models either shocked ISM or ejecta emission. As the computational times for the analysis is long, we restrict the analysis to the quickest models to run e.g. a thermal component with a simple power-law component to represent the non-thermal emission and a \texttt{nei} model for the shocked ISM. The ejecta is still modeled by a single \texttt{vnei} component.

\subsubsection{Non-thermal + thermal ISM emission}
\label{subsec:non_thermal_thermal_ISM_emission}

\begin{figure*}
    \centering
    \plotone{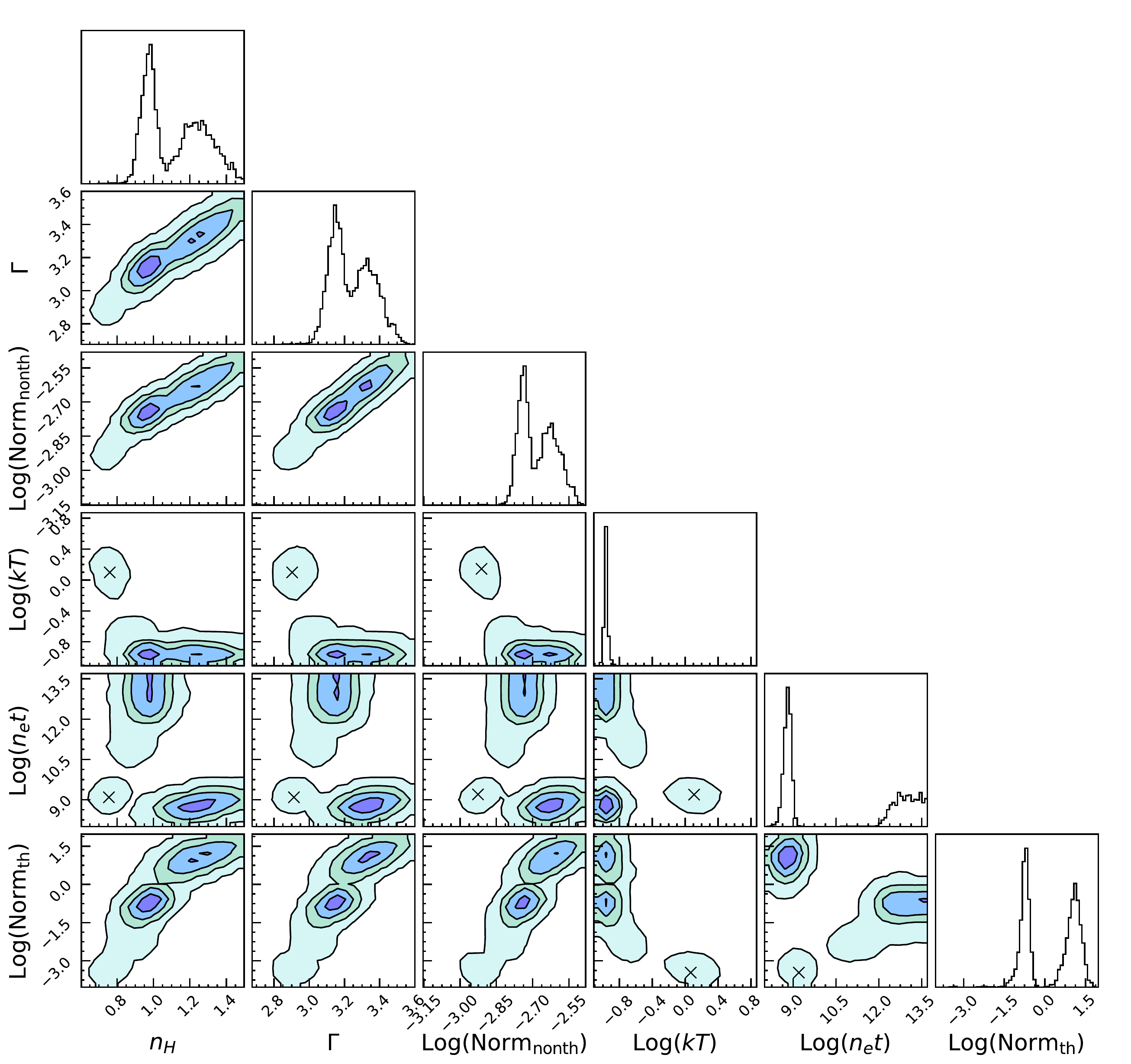}
    \caption{Posterior likelihood distribution of the \texttt{TBabs(powerlaw+nei)}~model on Region~2. The distributions have been sampled with the \texttt{BXA} package. Log-uniform priors have been set for all parameters, with the exception of the power-law photon index $\Gamma$ and the Hydrogen absorption $N_H$, which have received linear priors. The temperature $kT_{\rm e}$ is given in keV, the normalization of the power-law in keV$^{-1}$ cm$^2$ s$^{-1}$, the ionization timescale $n_et$ in s cm$^{-3}$ and the normalization of the thermal model in cm$^{-3}$. The top plot of each column shows the individual parameter histograms. The contours correspond (from darker to lighter blue) to $1\sigma$, $2\sigma$, $3\sigma$ and $5\sigma$ significance levels. The likelihood distribution features a complex multimodal shape which appears to be dominated by two main solutions. The areas marked with a black cross in the $n_et$ and $kT_{\rm e}$ boxes correspond roughly to the new boundaries for these parameters when enforcing self-consistency for the NEI model (see text in Section~\ref{subsec:non_thermal_thermal_ISM_emission} for details). This plot has been realized with the \texttt{corner} Python package \citep{Foreman2016}.}
    \label{fig:posterior_large_pownei}
\end{figure*}


\begin{figure*}
    \plottwo{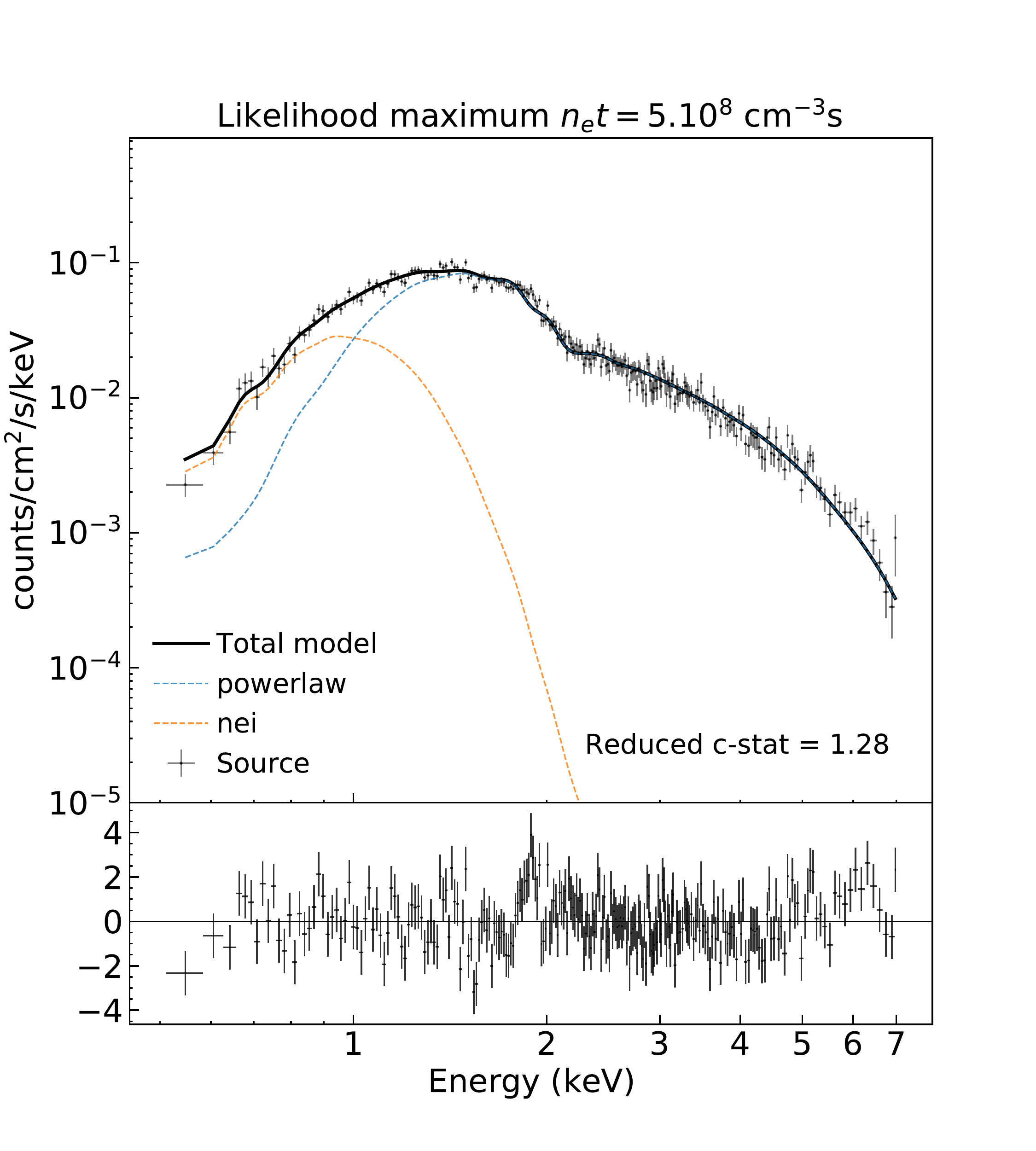}{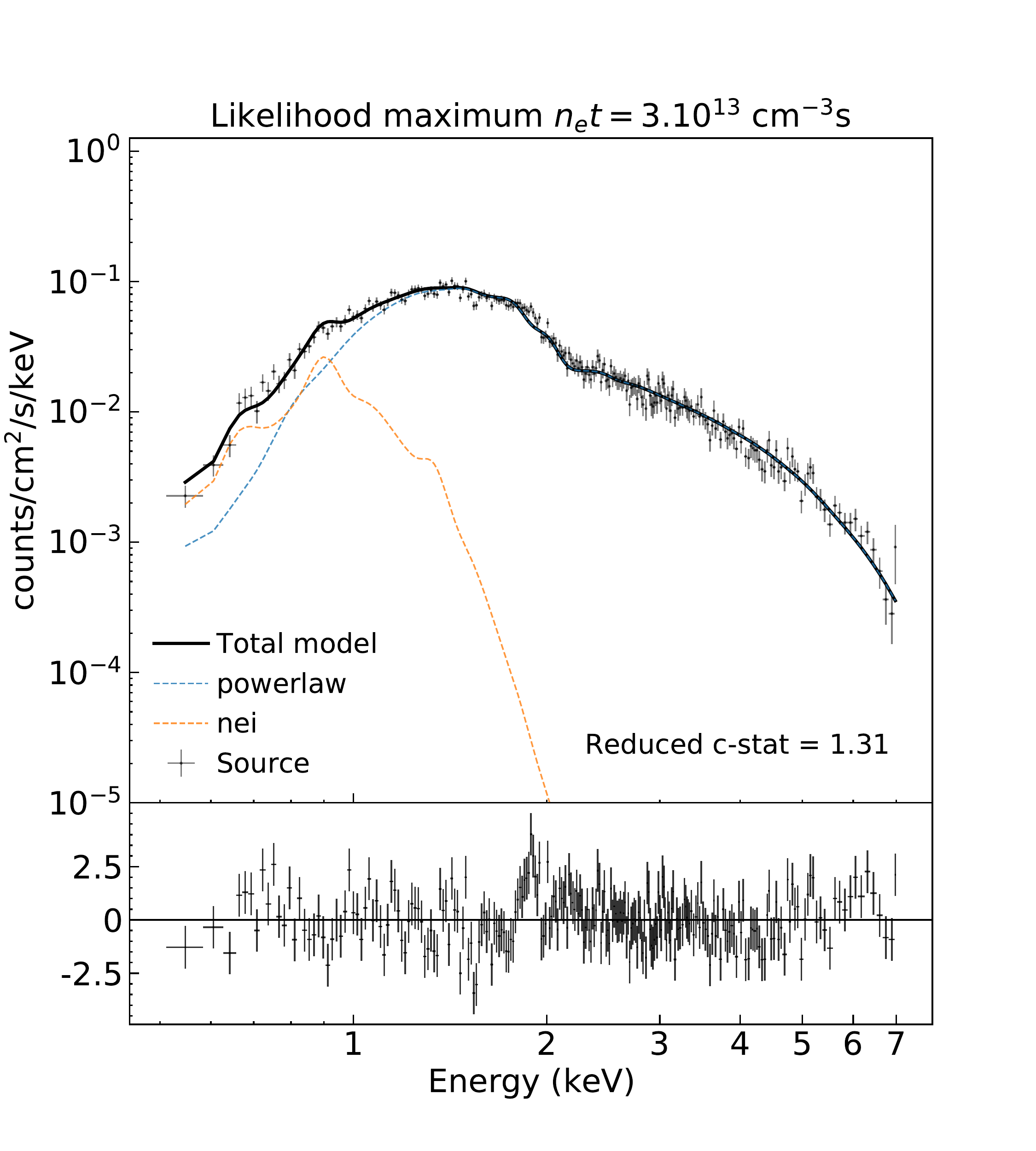}
   \caption{Left: best fit model and residuals associated to the likelihood maximum centered on $n_et = 5.10^8$~cm$^{-3}$s, for the \texttt{Tbabs(powerlaw+nei)} model over Region~2's spectrum. Right: same as left but the likelihood maximum is centered on  $n_et = 3.10^{13}$~cm$^{-3}$s.}
    \label{fig:stats_maxima}
\end{figure*}

A \texttt{Tbabs(powerlaw+nei)} model was used to model the synchrotron and ISM emissions.
We sampled the posterior distribution for all regions with the same priors as for the single component models, with the exception of the temperature which is assigned a log-uniform prior to speed up computation. We took Region~2 as main example and display the resulting posteriors in the corner plot in Figure~\ref{fig:posterior_large_pownei}. Note that the posteriors for this model are similar to first order to those for the other regions.

The likelihood distribution displays a complex shape, with several peaks of which two modes strongly dominate. These two modes cover values of $n_et$ centered on $\sim5.10^{8}$~cm$^{-3}$s and on $\sim3.10^{13}$~cm$^{-3}$s respectively, and both indicate small $kT_{\rm e}$ values of $\sim 0.1$~keV. The two corresponding best fits are shown in Figure~\ref{fig:stats_maxima} with their residuals and reduced \textit{C--stat} values (denoted as $C_r$). The power-law component dominates the emission, while the peak of the thermal emission lies below an energy of 1 keV. While both $C_r$ ($\sim1.3$) indicate the fitted models to be statistically acceptable, they are barely able to reproduce faint spectral line features present in the spectra. Most notably, the signature of the Si emission line at 2~keV can be seen in the residuals. This indicates difficulties for this two-component non-thermal+ISM model to fully reproduce the observed spectra.

Additionally, an $n_{\rm e}t$ as  high as $3.10^{13}$~cm$^{-3}$s is unlikely to originate from shocked ISM in the vicinity of the blast wave. The maximum ionization timescale is related to the shock velocity and the width of the emission region as
\begin{align}\label{eq:net}
  n_{\rm e}t\approx &4\times n_0\Delta t 
  = 4n_0 \frac{\Delta r}{V_{\rm s}/4}\\\nonumber
  =& 1.4\times 10^{10} n_0 \left(\frac{\Delta r}{0.1~{\rm pc}}\right)\left(\frac{V_{\rm s}}{3500~{\rm km\,s^{-1}}}\right)^{-1}~{\rm cm^{-3}s},
\end{align} 

with $n_0$ the pre-shock density, $\Delta r$ the width of the extraction region, and assuming a shock compression factor of four.
Assuming a distance $D=2.3$~kpc for Tycho's SNR, the width of our spatial regions is $\Delta r=$6\arcsec $\simeq$ 0.087~pc. 
Taking a shock velocity 
$V_{\rm s} = 3328$~km.s$^{-1}$
(see Table~\ref{tab:williams_temperatures}), 
an  $n_{\rm e}t$ as high as $3\cdot 10^{13}~{\rm cm^{-3}s}$
would imply an improbably high pre-shock electron density of $>10^3{\rm cm^{-3}s}$.
This is orders of magnitudes above the measured
electron densities reported in the literature,
which range from $\sim0.1$ to $\sim1.5$~cm$^{-3}$\citep{Chiotellis2013}.
 Instead, for the extraction width used,  $V_{\rm s} = 3328~{\rm km\ s^{-1}}$ and $n_0=0.3~{\rm cm^{-3}}$
 we expect $n_{\rm e}t\approx 4\times 10^{9}~{\rm cm^{-3}s}$. 

The low $n_{\rm e}t$ values ($\sim5.10^8$~cm$^{-3}$s) given by the other mode correspond to reasonable post-shock electron densities ($\sim0.2$~cm$^{-3}$). However, the thermal normalization is also sensitive to electron density, and to emission volume. One can compute the electron density by assuming a depth along the line of sight for the spatial regions we are analysing such as $n_e^2\approx \mathrm{norm} \times 4\pi  VD^210^{14}$. Assuming the emission volume $V$ to be a box with a depth equals of same size as the length of our analyzed regions (90\arcsec~in sky plane) gives a post-shock electron density of $\sim$~9.5~cm$^{-3}$, which is two orders of magnitude higher than the value obtained from the ionization timescales. Varying the depth of the volume does not change significantly this result, unless it is over order of magnitudes compared to the length of the regions (a depth at least 100 times the length of the region is needed to bring the density back to consistent values). This would imply a strongly anisotropic shape for the remnant with an over elongated morphology along line of sight, which is not plausible. As this mode is not self-consistent, we discarded it too.

The inconsistencies of the \texttt{XSPEC} NEI model led \citet{Cassam2007} to use a self-consistent NEI model by first assuming a postshock electron density and then computing the corresponding emission measure and ionization timescale. We took a similar approach by modifying the boundaries  of these two parameters based on assumed electron densities values, i.e. we imposed additional priors. We chose an upper bound value of 10~cm$^{-3}$ for the postshock electron density, giving upper bounds of $\sim5.10^{-3}$~cm$^{-3}$ and $\sim5.10^{9}$~cm$^{-3}$s for the thermal normalization and ionization timescale respectively. We then sampled the likelihood posterior distribution of the \texttt{TBabs(powerlaw+nei)} model with these new constraints. This sampled region in the parameter space is actually part of the $5\sigma$ contours drawn in Figure~\ref{fig:posterior_large_pownei}, and corresponds approximately to the third isolated region visible in several boxes and marked by a small black cross. The corresponding best fits are displayed in Figure~\ref{fig:powvneifree} and the derived parameter values are displayed in Figure~\ref{fig:plot_bicomponent_values_3by2}. The resulting electron temperature values are ranging between 2 and 2.6 keV for all regions but Region~2 that displays a lower value of 1.3 keV. The hydrogen column density varies from 0.77 to 1.1$\times10^{22}$cm$^{-2}$, and shows a similar trend as the electron temperature, with the minimum value given by Region~2. The ionization timescale values range between $1-5\times 10^{9}$~cm$^{-3}$s. The photon indexes values range between 2.9 and 3.2. Since the power-law normalization values are very similar to the single power-law model already shown in Figure~\ref{fig:plot_single_component_values_3by2}, we do not display them here.


\subsubsection{Non-thermal + ejecta emission}
\label{subsec:non_thermal_ejecta_emission}

\begin{figure*}
\centering
\plotone{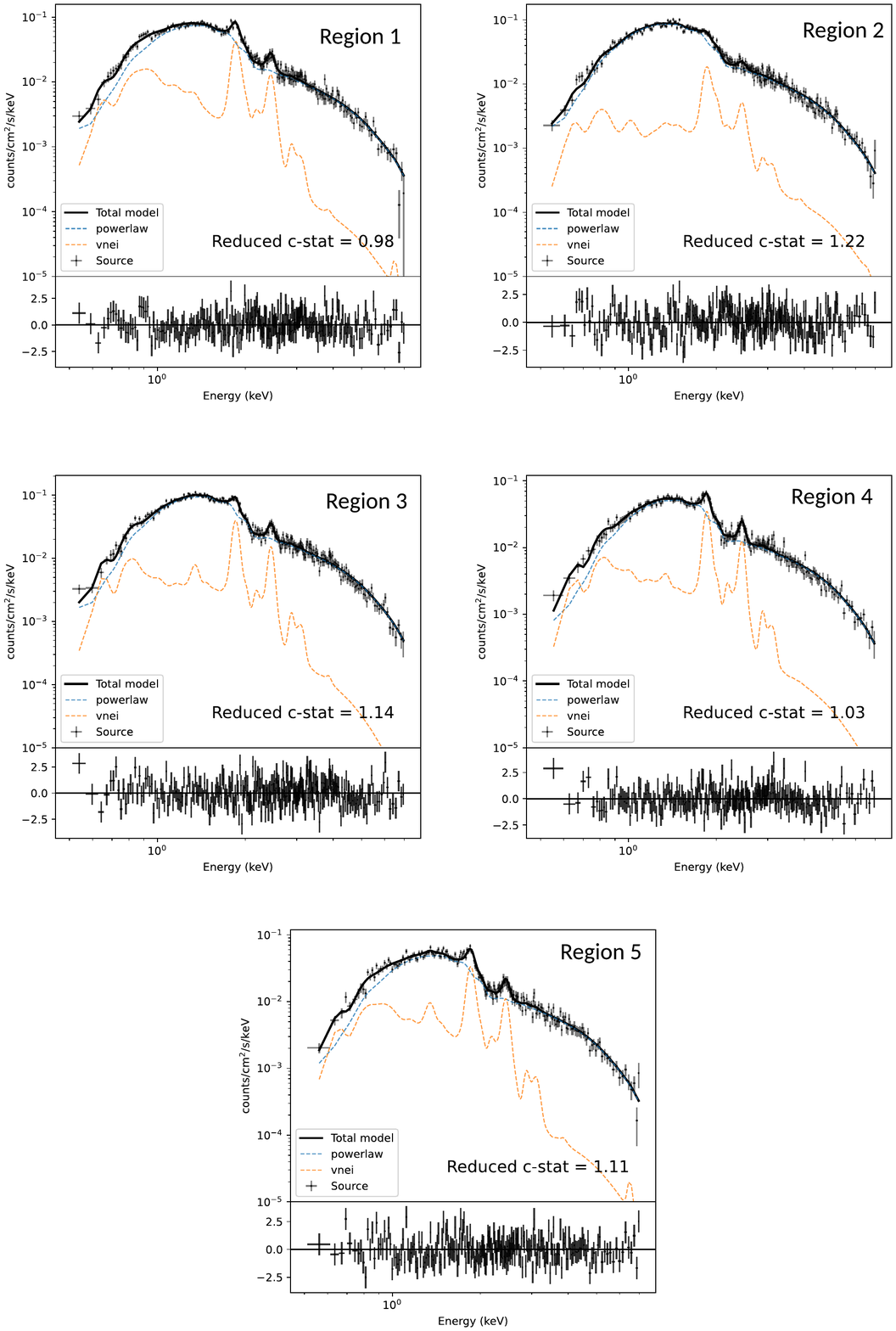}
\caption{Best bayesian fit for the \texttt{Tbabs(powerlaw+vnei)} model for all five regions. This model is the best two-components model to reproduce the spectra, according to bayesian evidences (see Table~\ref{tab:regions_logz}).}
\label{fig:powvneifree}
\end{figure*}

A \texttt{Tbabs(powerlaw+vnei)} model was used to model the synchrotron and ejecta emissions. A problem with the \texttt{vnei} model implemented in \texttt{XSPEC} is that the continuum is computed from hydrogen abundance, which is not relevant in the case of ejecta-dominated spectra. A way to bypass this issue as well as the emission measure / metal abundance degeneracy is to set the heavy element abundances to high values. This way, the thermal continuum emission is dominated by the metal-rich ejecta and the continuum due to hydrogen is negligible---see \citet{Greco2020} for more details. The price of this approach is an uncertainty on the absolute value of abundances in the regions selected which is acceptable here as our analysis is focused on the shocked ISM emission rather than on the ejecta one.

The same parameters ranges as for the corresponding single component model were set for both the components, with the exception of abundances and normalization for the \texttt{vnei} due to the abundance/emission measure degeneracy in \texttt{XSPEC} thermal models. The abundances were set to vary between higher values (between $10^{2}$ and $10^{4}$), and the thermal normalization to lower values (between $10^{-10}$ and $10^{-7}$) in order to properly reproduce the ejecta-dominated scenario. Note that larger ranges for the abundances were not possible due to computation time issues.

The resulting posteriors are monomodal for all parameters, with the exception of abundances (similarly to the single \texttt{vnei} component) and of the electron temperature. Depending on the region, most element posterior distributions are either flat or falling on parameter boundaries, with the exception of the O, S and Si elements which are in average well constrained. For all regions the electron temperature posterior distribution was flat over the whole parameter space.

\begin{figure*}
\centering
\plotone{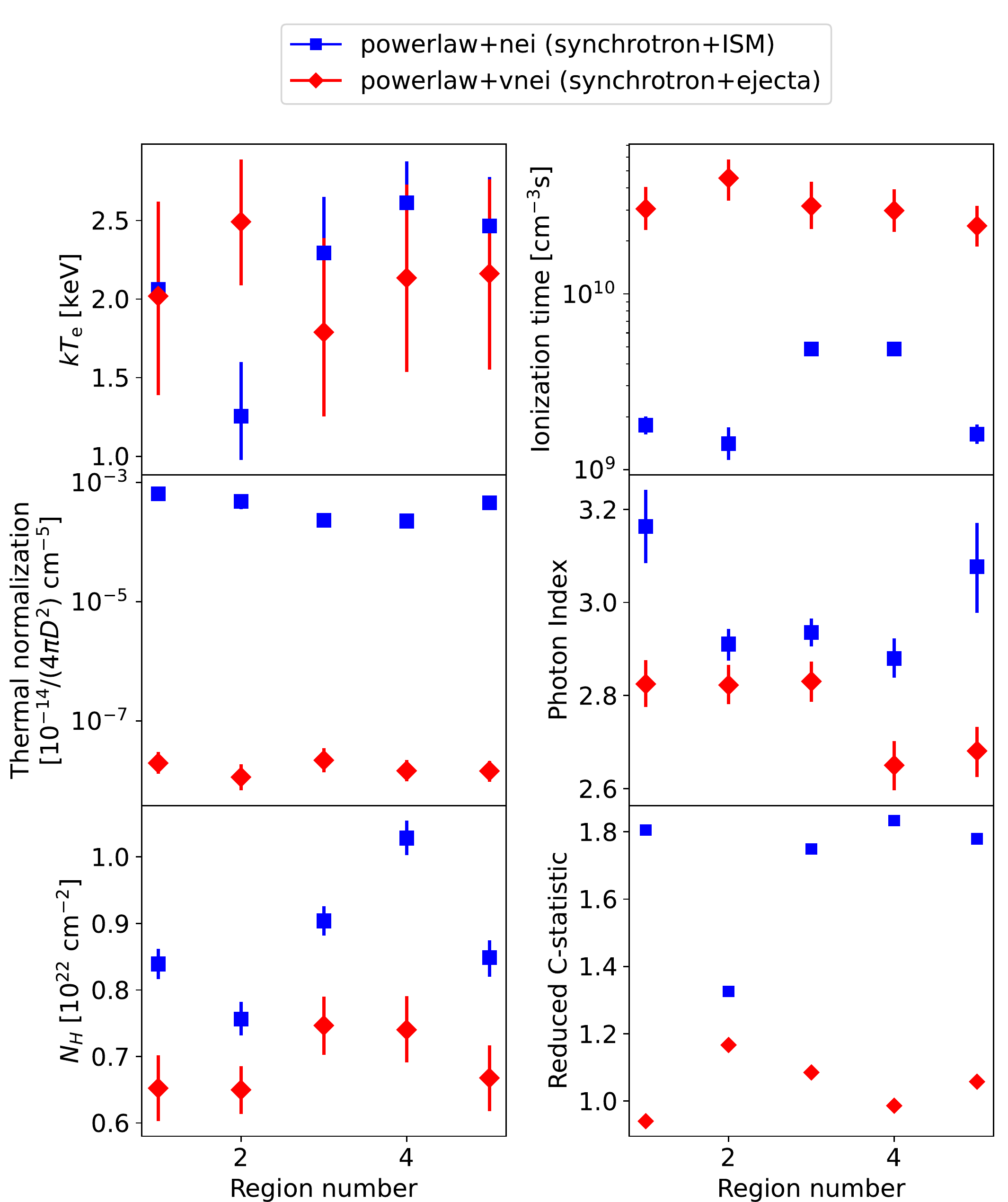}
\caption{Best Bayesian parameter values with 1-$\sigma$-equivalent quantile error bars for two-components models models in each of the five regions. Only the main parameters and the \textit{C-stat} are shown. The physical component for each model is indicated in the legend. The thermal normalizations are so different because of our method to take care of the abundance/emission measure degeneracy \citep[see text and ][ for details]{Greco2020}. Some error bars are smaller than the corresponding data points and are therefore not visible.}
\label{fig:plot_bicomponent_values_3by2}
\end{figure*}

The best Bayesian fits for all regions are displayed in Figure~\ref{fig:powvneifree}, with their goodness-of-fit. The derived parameter values are shown in Figure~\ref{fig:plot_bicomponent_values_3by2}. The Si abundance best-fit values for all regions are of the order of $10^{4}$. The abundance ratio of other elements by Si can be found Table~\ref{tab:ab_ratios}. While the mean normalization of the ejecta component is much lower than the non-thermal component, the modeling of the emission lines allows to fit some crucial thermal features such as the Si line at 2~keV, on the contrary to the non-thermal+ISM model (see residuals in Figures~\ref{fig:posterior_large_pownei} and \ref{fig:stats_maxima}).

\subsubsection{Comparison of two-components models}

In the same way as Section~\ref{subsub:comparison_single_component_models}, we use the Bayes factors derived by \texttt{BXA} to compare the two-components models. The Bayes factors are displayed in Table~\ref{tab:regions_logz}, alongside a selection of single component models and of a three component model that are discussed later in this work.

When comparing two-components non-thermal+ISM models to the best single-component models, one can clearly see that in all regions but region~2, the single \texttt{vnei} performs better than the non-thermal+ISM models. This is due to ejecta lines not being reproduced either by the power law or by the ISM model. Due to the ability of the ejecta component to reproduce these faint spectral line features in the model on top of the power-law, the non-thermal+ejecta gives a better representation of the shock spectra than the non-thermal+ISM and single component models. This is clearly visible in the Bayes factors displayed in Table~\ref{tab:regions_logz} and enhances the fact that accurately modeling the thermal features present in the spectra deserves considerable attention when studying non thermal X-ray emission from shocks, even if they are synchrotron dominated. This also shows that the ejecta emission is the dominant thermal contribution to the spectra.

However, there are still spectral features not correctly modeled by the non-thermal+ejecta model and visible in the residuals (see Figure~\ref{fig:powvneifree}), especially at lower energy ($<2$~keV). This still leaves room for an improved representation of the spectra, using a third component.

\begin{table}

    \caption{Ejecta abundance ratios derived from the two-component non-thermal+ejecta model (\texttt{Tbabs(powerlaw+vnei)}.}
    \centering
    \begin{tabular}{c c c c c c}
        \hline\hline
        & Region 1 & Region 2 & Region 3 & Region 4 & Region 5 \\
O/Si & 1.391 & 0.589 & 0.458 & 0.831 & 0.489 \\
Ne/Si & 0.046 & 0.043 & 0.047 & 0.055 & 0.039 \\
Mg/Si & 0.053 & 0.052 & 0.066 & 0.084 & 0.040 \\
S/Si  & 0.540 & 1.409 & 1.474 & 1.794 & 1.236 \\
Ar/Si & 0.064 & 0.080 & 0.077 & 0.096 & 0.078 \\
Ca/Si & 0.065 & 0.067 & 0.074 & 0.095 & 0.072 \\
Fe/Si & 0.046 & 0.049 & 0.111 & 0.088 & 0.143 \\
        \hline
                \\
            \end{tabular}

            \label{tab:ab_ratios}

        \end{table}

\begin{table*}

\caption{Bayes factors for a selection of single, two-components and three component models in each of the five regions.}
\label{tab:regions_logz}
\centering
\resizebox{\textwidth}{!}{
\begin{tabular}{l l c c c c c c}
\hline\hline\\
& & \multicolumn{6}{c}{Log(z) - Log(z)$_\mathrm{max}$} \\\\\cline{3 - 8}\\
& Model & Region 1 & Region 2 & Region 3 & Region 4 & Region 5 & Average \\\\
\hline\\
\multirow{2}{*}{Single non-thermal} 
& \texttt{TBabs(sqrtcutoffpl)} & -231.34  & -45.14  & -146.64  & -211.74  & -200.10  & -166.84 \\ 
& \texttt{TBabs(vnei)} & -54.42  & -51.14  & -91.86  & -58.91  & -45.28  & -60.17 \\ 
\\
\hline
\\
\multirow{1}{*}{Non-thermal + thermal} 
& \texttt{TBabs(powerlaw+nei)} & -196.61  & -40.31  & -150.37  & -192.61  & -161.36  & -148.10 \\ 
\\
\hline
\\
\multirow{1}{*}{Non-thermal + ejecta} 
& \texttt{TBabs(powerlaw+vnei)} & -12.32  & -9.48  & -5.57  & -7.86  & -7.98  & -8.49 \\ 
\\
\hline
\\
\multirow{2}{*}{Non-thermal + ejecta + ISM} 
& \texttt{TBabs(powerlaw+nei+vnei)} & -0.18  & \cellcolor{green}0.00  & -1.37  & \cellcolor{green}0.00  & -0.20  & -0.20 \\ 
& \texttt{TBabs(powerlaw+nei+vnei)}\tablenotemark{a} & \cellcolor{green}0.00  & -0.06  & \cellcolor{green}0.00  & -0.68  & \cellcolor{green}0.00  & \cellcolor{green}0.00 \\ 
\\
\hline
\end{tabular}
}
\tablenotetext{}{Notes. For each region (column) the Bayesian evidence of every model is normalized to the best one in that region (column). In the last right column is shown the Bayesian evidence for each model averaged over all regions, normalized by the best average evidence. The best model for each region has a value of zero and is highlighted in green.}
\tablenotetext{a}{Slightly wider priors have been used for this model to get better constraints on the parameters. More information is given in the text (see Section~\ref{sec:three_components}).}

\end{table*}

\subsection{Three component models}
\label{sec:three_components}

In this section we combine two thermal components alongside a non-thermal one. Computation time issues are even more prevalent here, forcing us to make prior assumptions about the models and strongly limiting the variety of models we can use. First, we use the simplest three component model \texttt{TBabs(powerlaw+nei+vnei)} to analyse all regions and compare it to single component and two-components models. Then we perform an extended analysis of Region~2 only, with non-thermal model variants.

\subsubsection{Non-thermal+ejecta+ISM}

\begin{figure*}
\centering
\plotone{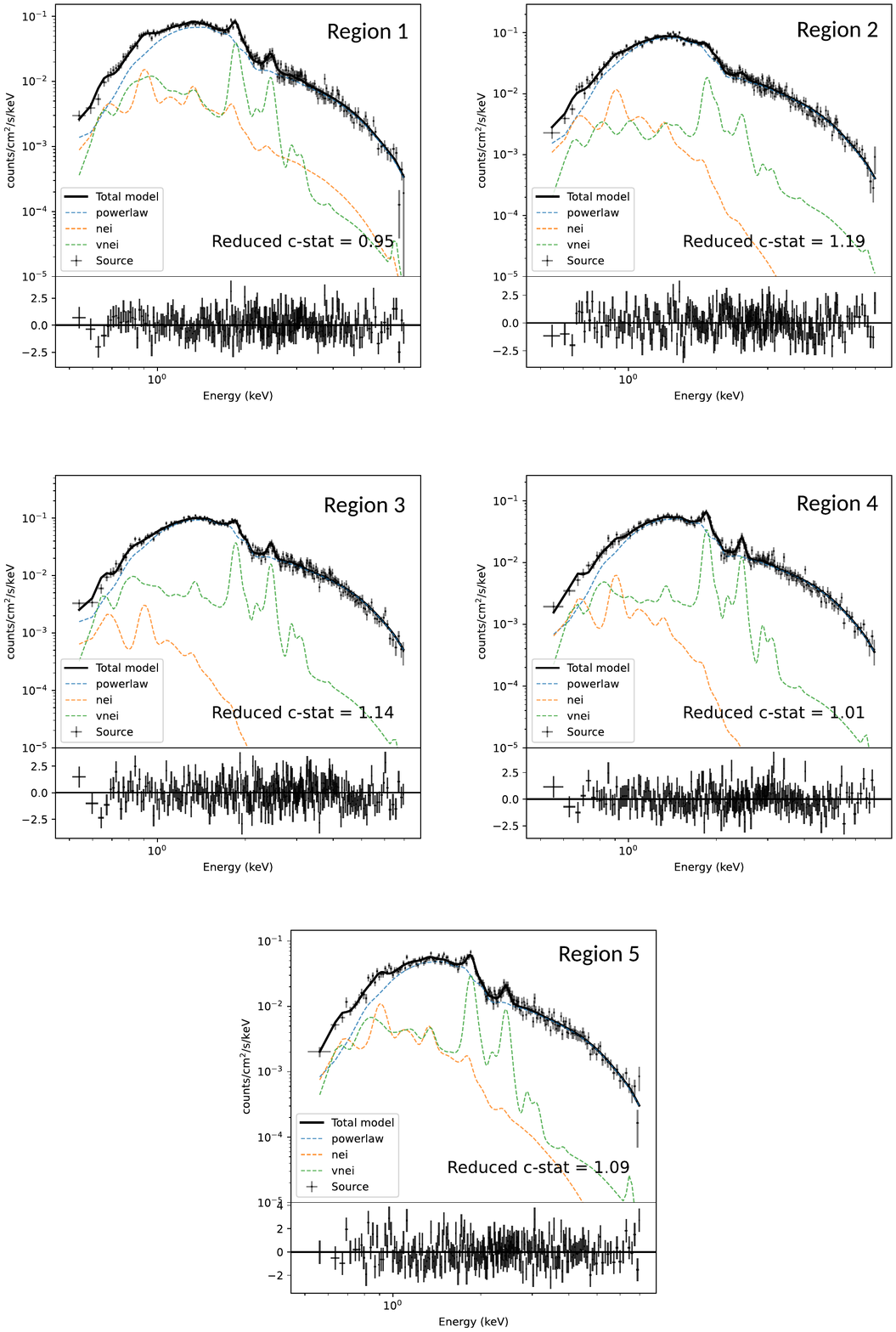}
\caption{Best bayesian fit for the \texttt{Tbabs(powerlaw+nei+vnei)} model for all five regions. The \texttt{nei} component models the shocked ISM emission, and the \texttt{vnei} component models the ejecta emission. The corresponding posterior distributions can be found in Appendix~\ref{app:posterior_all_regions}.}
\label{fig:powpshockvnei}
\end{figure*}


\begin{figure*}
\centering
\plotone{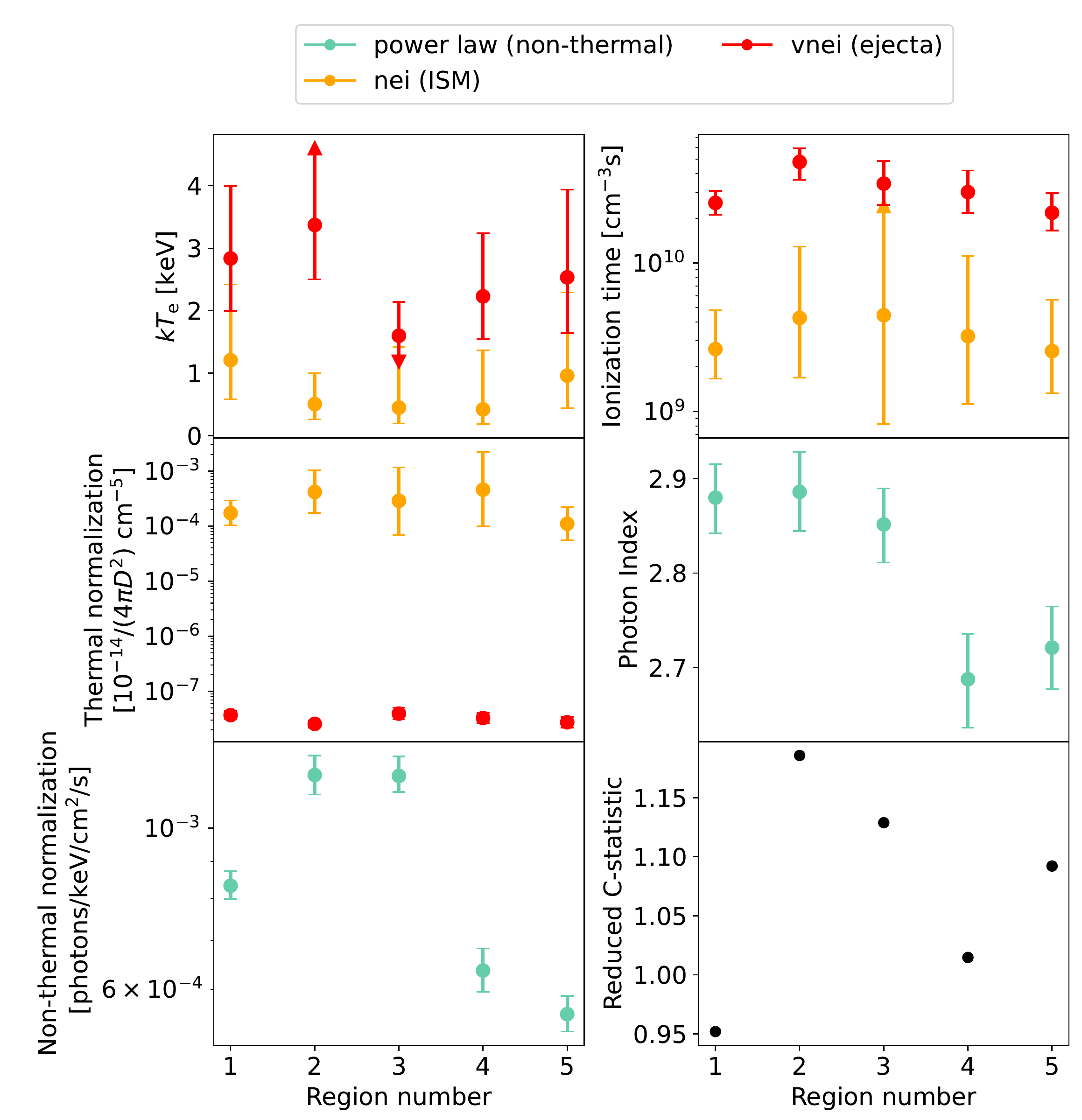}
\caption{Best Bayesian parameter values with 1-$\sigma$-equivalent quantile error bars for \texttt{TBabs(powerlaw+nei+vnei)} models in each of the five regions. Note that in this plots, the different colors correspond to model components, and not to models by themselves. Only the main parameters and the \textit{C-stat} are shown. The physical component for each component is indicated in the legend. The thermal normalizations are so different because of our method to take care of the abundance/emission measure degeneracy \citep[see text and ][ for details]{Greco2020}. Some error bars are smaller than the corresponding data points and are therefore not visible.}
\label{fig:plot_tricomponent_values_3by2}
\end{figure*}


\begin{figure}
\centering
\plotone{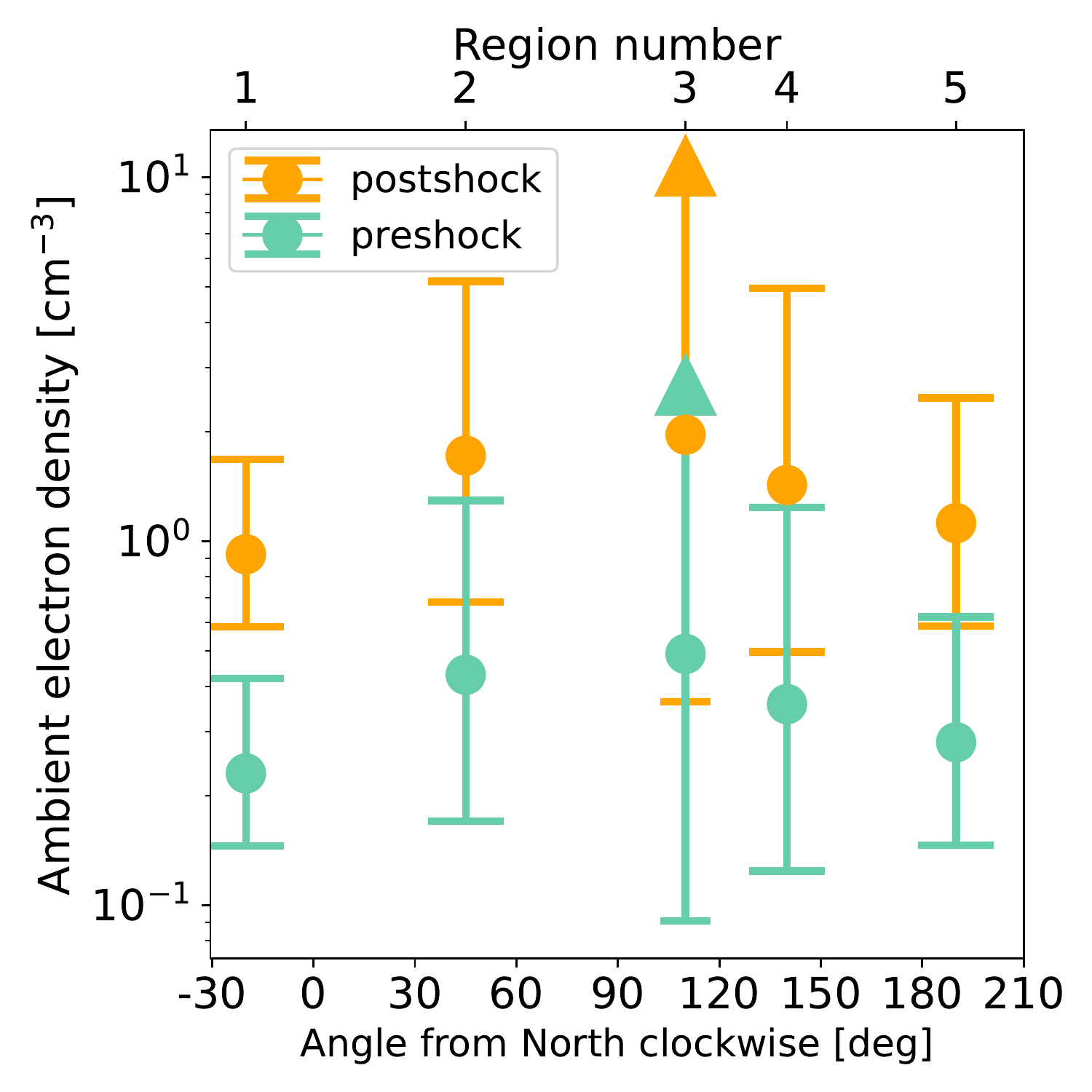}
\caption{Postshock electron densities using $n_et$ values from the ISM component of the \texttt{TBabs(powerlaw+nei+vnei)} model, shock velocities from \citet[][see Table~\ref{tab:williams_temperatures}]{Williams2016}. The ambient (preshock) electron densities are computed with a compression ratio of 4. Horizontal axes show the region numbers and the region azimuthal angles in degrees clockwise from the north.}
\label{fig:plot_ne}
\end{figure}


A \texttt{TBabs(powerlaw+nei+vnei)} was used to model respectively the synchrotron, ISM and ejecta emission. For the ejecta priors, the $kT_{\rm e}$ was first allowed to vary between 1 and 5 keV. The ionization timescale was still left free and was allowed to vary over the whole default \texttt{XSPEC} parameter space. Leaving abundances free to vary were resulting in computation times so long it was not converging in realistic times for this work. Therefore, we chose to freeze abundances using prior information.
The Si abundance was frozen to $10^4$, and the other abundances were set (and also frozen) accordingly to the ratios previously computed (see Table~\ref{tab:ab_ratios} and Section~\ref{subsec:non_thermal_ejecta_emission}). The priors for the second thermal component (ISM) are the same as described in Section~\ref{subsec:non_thermal_thermal_ISM_emission}, most notably with the self-consistency constraints.

The parameter posterior distributions are for the most part monomodal, although not necessarily gaussian-like and are, in a few cases, poorly constrained. Most notably the tail of the 1-$\sigma$ contours of the ISM ionization timescale is cut by the upper boundary for Regions 2, 3, 4 and 5. The ISM electron temperature posterior is similarly cut in Regions 3, 4 and 5. This might be an issue to estimate upper limits on these parameters. In order to properly constrain these parameters, we relaxed slightly the priors on the upper boundary. We relaxed the ISM electron temperature upper limit from 3 to 5~keV for all regions. We also relaxed the ISM ionization timescale upper limit from $5\times10^9$ to $9\times10^9$~cm$^{-3}$s in Regions 4 and 5, and to $3.5\times10^{10}$~cm$^{-3}$ for Regions 2 and 3. These upper ranges are well above the expected values of $n_{\rm e}t$ of $4\times 10^9~{\rm cm^{-3}s}$, see Section 3.5.1, Eq.~\ref{eq:net}. These wider priors allow to correctly constrain the 1-$\sigma$ contours for the temperature in all regions, and to constrain the ionization time in Regions 2, 4 and 5. However the ISM ionization time remains unconstrained in Region~3. Given that the estimated pre-shock densities are of the order of 0.2~cm$^{-3}$, with a maximum possibility of $n_0=1.5~{\rm cm^{-3}}$ -- implying $n_{\rm e}t<2\times 1.8\times 10^{10}~{\rm cm^{-3}s}$ -- setting even higher upper boundary would make the model physically implausible. 

Our analysis, therefore, only provides a lower constraint for this specific parameter in Region~3. We stress that modifying the priors here has a very low impact on Bayesian evidence values, as shown in Table~\ref{tab:regions_logz}. The resulting posterior distributions are shown in Appendix~\ref{app:posterior_all_regions}). We show the best fits for the \texttt{Tbabs(powerlaw+nei+vnei)} model in Figure~\ref{fig:powpshockvnei} with their associated $C_r$. The best fit parameter values are shown in Figure~\ref{fig:plot_tricomponent_values_3by2}.

As seen in Figure~\ref{fig:plot_tricomponent_values_3by2}, the mean electron temperature over all our shock regions for the ISM thermal component is $kT_{\mathrm{e,ISM}}=0.96^{+1.33}_{-0.51}$~keV, which is lower than the mean ejecta temperature $kT_{\mathrm{e, ejecta}}=2.53^{+1.40}_{-0.89}$~keV. The average best-fit values found for the ISM ionization timescale in all regions (excluding Region 3 as it only provide a lower limit) is $(n_{\rm e}t)_{\rm ISM}=2.55^{+0.5}_{-1.22}\times10^9$~cm$^{-3}$s, which is an order of magnitude lower than the average ejecta ionization timescale $(n_{\rm e}t)_{\rm ejecta}=2.17^{+0.0.76}_{-0.52}\times10^{10}$~cm$^{-3}$s. We computed the postshock electron densities using shock velocities from Table~\ref{tab:williams_temperatures}, and the ambient densities for each region using a compression ratio of 4 (see Section~\ref{subsec:non_thermal_thermal_ISM_emission} for the detailed methodology). The preshock and postshock electron densities are displayed in Figure~\ref{fig:plot_ne}, and show the same trend as $n_et$. We find an average (excluding Region 3) ambient electron density around $n_{\rm e}=0.32^{+0.23}_{-0.15}$~cm$^{-3}$.

While the statistics are low with only five points, it is still interesting to compare the variations of shocked ISM properties along the edge of the SNR.
The hydrogen column density was left free to vary during our analysis. As the shocked ISM emission resides mainly at low energy, strong hydrogen column density differences between our regions could bias our comparison. This is not the case here as we find only moderately higher $N_H$ best-fit values for regions~3 and 4 with an average \ama{$N_H=0.78^{+0.04}_{-0.04}10^{22}$~cm$^{-2}$} than for the other regions with an average \ama{$N_H=0.71^{+0.04}_{-0.04}10^{22}$~cm$^{-2}$}.
We investigated the presence of differences between regions 1 and 2 (northern area of the remnant) and regions 4 and 5 (southern area, we exclude Region~3 here because it only provides lower limit on the density), as a ambient medium density gradient has been found in the literature around Tycho SNR \citep{Williams2013}. \ama{Northern regions have an average electron density $n_{\mathrm{e}}=0.33^{+0.19}_{-0.15}$~cm$^{-3}$, while southern regions have $n_{\mathrm{e}}=0.31^{+0.29}_{-0.13}$~cm$^{-3}$. This difference is not strong enough to confirm the presence of a density gradient, especially considering the large error bars.}
However the density values found are consistent with values from literature, whether they are from previous shock X-ray spectral studies \citep[$\leq0.7$~cm$^{-3}$;][]{Cassam2007}, X-ray expansion measurements \citep[$\sim0.2$~cm$^{-3}$;][]{katsuda2010} or infrared flux measurements \citep[e.g. $\sim0.1-0.2$~cm$^{-3}$]{Williams2013}. 

On the other hand, there is no evidence for a particular correlation between the properties of the shocked ISM and the properties of the non-thermal synchrotron emission.

\subsubsection{Comparison with previous models}
\label{subsec:comparison_model_tri}
The resulting Bayesian evidence values are displayed alongside previous models in Table~\ref{tab:regions_logz}. The three component models give significantly better representations of the spectra than previous models. The associated \DL~is larger than the limit of 2 set by Jeffrey's scale, and way beyond uncertainties due to numerical errors. This trend is confirmed by other metrics such as the AIC (see Table~\ref{tab:aic}) and the $\Delta C$ (see \ref{tab:deltac}), but is in disagreement with the BIC (see \ref{tab:bic}). This is purely due to the abundances being frozen in the three component model contrary to being free to vary in the two-components model, creating a difference in number of parameters skewing the BIC numerical values by a large amount. In addition to the low electron temperatures and ionization times, this supports the fact that this second thermal component is actually a signature of faint shocked ISM. This signature is present in all regions, although poorly constrained in Regions~3, 4 and 5, as shown by the posterior distributions of the ISM parameters in Figures~\ref{fig:posterior_powneivnei_3}, \ref{fig:posterior_powneivnei_4} and \ref{fig:posterior_powneivnei_5}).

\subsubsection{Extended analysis of Region~2}

\begin{table*}
\caption{Bayes factors and other goodness-of-fit metrics for a selection of models for Region~2.}
\label{tab:logz_table_reg2}
\centering
\resizebox{\textwidth}{!}{
\begin{tabular}{l l c c c c}
\hline\hline\\
& Model & $\log(z) - \log(z)_\mathrm{max}$ & $\Delta C$ & $AIC - AIC_{\mathrm{min}}$ & $BIC - BIC_{\mathrm{min}}$ \\\\
\hline\\ 
\multirow{2}{*}{Single non-thermal} 
& \texttt{TBabs(powerlaw)} & -74.37  & 148.73  & 136.73  & 196.83 \\ 
& \texttt{TBabs(sqrtcutoffpl)} & -45.14  & 90.28  & 80.28  & 132.28 \\ 
\\
\hline
\\
\multirow{2}{*}{Non-thermal + ISM} 
& \texttt{TBabs(powerlaw+nei)} & -40.31  & 80.62  & 74.62  & 110.43 \\ 
& \texttt{TBabs(sqrtcutoffpl+nei)} & -33.77  & 67.54  & 65.54  & 85.16 \\ 
\\
\hline
\\
\multirow{2}{*}{Non-thermal + ejecta} 
& \texttt{TBabs(powerlaw+vnei)} & -9.48  & 18.96  & 28.96  & \cellcolor{green}0.00 \\ 
& \texttt{TBabs(sqrtcutoffpl+vnei)} & -14.08  & 28.15  & 22.15  & 57.96 \\ 
\\
\hline
\\
\multirow{2}{*}{Non-thermal + ejecta + ISM} 
& \texttt{TBabs(powerlaw+nei+vnei)} & \cellcolor{green}0.00  & \cellcolor{green}0.00  & \cellcolor{green}0.00  & 11.52 \\ 
& \texttt{TBabs(sqrtcutoffpl+nei+vnei)} & -0.22  & 0.43  & 0.43  & 11.95 \\ 
\\
\hline
\end{tabular}}
\tablenotetext{}{Notes. The models were chosen to show the differences between models with a cutoff non-thermal component and models with a regular power law. The complete version of this table with values for all 35 models can be found in Table~\ref{tab:logz_table_reg2_all}. The best model for each region has a value of zero and is highlighted in green.}
\end{table*}


\label{subsec:extended_analysis_of_region_2}

In the case of multiple component models, we so far restrained ourselves to the simplest phenomenological model for the non-thermal component, e.g. a power-law, and to the simplest thermal models for the shocked ISM (\texttt{nei)}. This was done to get a global picture across all five regions while still maintaining computation times in realistic time ranges. In order to test models with more variants, we now only use Region~2 as main analysis region in the following sections, increasing the number of models tested but diminishing the number of regions analyzed.

We sampled the composite (e.g. 'non-thermal + thermal', 'non-thermal+ejecta' and 'non-thermal+ejecta+ISM') model posterior distributions, with the same setups for the non-thermal components. The thermal components were given the exact same parameter boundaries than in Sections~\ref{sec:single_component_models} and \ref{sec:bicomponent_models}. The \texttt{BXA} sampler had difficulties with three component models featuring an \texttt{srcut} component. These models were the longest to sample, with weeks of computation for a single model running on 42 cores\footnote{Explaining why this part of our spectral analysis has been done over one region only.}. The other composite models with cutoff non-thermal component showed similar issues, although at slightly smaller extents. The corresponding Bayes factors are displayed for a subsets of the models alongside other metrics in Table~\ref{tab:logz_table_reg2}.

When looking at Table~\ref{tab:logz_table_reg2}, one can see that the 
single-component models with energy cutoff perform better than the single-component thermal models and regular power law. But a different picture appears when adding a component modeling shocked ISM or ejecta, where the presence of a cutoff gives either lower bayesian evidence (non-thermal+ejecta case) or no significant difference in bayesian evidence (non-thermal+ejecta+ISM case). The highest evidence is admittedly given by the \texttt{TBabs(powerlaw+npshock+vnei)} model, but other three component models have close evidence values (see Table~\ref{tab:logz_table_reg2_all} in appendix). As explained in Section~\ref{subsec:metrics}, these values are too close to each other to make definitive conclusions about which model is best.
%

We will not go into too many details about posterior distribution shapes here, but still highlight a few interesting points. The posteriors of three component models with a cutoff are for most case monomodal, so we do not display them.  But the energy cutoff parameter for these models features an elongated 'banana' shape and is almost always cut by upper boundaries. This behaviour did not change when we tried to increase the upper boundary value, which makes the cutoff energy of these models poorly constrained. While the Bayes factors do not indicate strongly whether a cutoff is necessary, or not, to describe the spectrum of Region~2, the poor constraints on this additional parameter indicates there is no need for the additional parameter. Applying Occam's rule to two models performing equally at describing the data, the simpler model should be chosen. In that case, it appears that a cutoff for the non-thermal component is unnecessary for the three component class of models, as it does not improve significantly the model performances at reproducing the observed shock spectrum. Therefore the presences of thermal components in the model and of an energy cutoff for the non-thermal component appear anti-correlated in Region~2. This casts some doubt on the cutoff energies obtained in previous work, which may have been affected by ignoring the presence of a (weak) thermal component. This should be further investigated, also in relation to the unexpected relation between cutoff energy and shock velocity reported in \citet{Lopez2015}.

\subsection{Best Bayesian selected model}
\label{subsec:best_bayesian_selected_model}

We added different cutoff models to our analysis in the previous section, increasing the total number of models applied to Region~2 spectrum to 35 (see Table~\ref{tab:logz_table_reg2_all}). The main point we want to highlight here is that adding the cutoff models did not change much the results found in Sections~\ref{sec:bicomponent_models} and \ref{sec:three_components} for Region~2. Bayes factors and other metrics indicate that three component models that take into account ejecta contamination should still be preferred over the other simpler models. This is an additional argument in favor of the presence of X-ray emission from shocked ISM behind the blast wave for not only Region~2, but for all regions. X-ray upper limits were previously provided by \citet{Cassam2007} on shocked ISM, based on the lack of evidence for thermal emission. While our findings  provide bayesian evidence for shocked ISM emission and should be confirmed later by spectroscopic data with an enhanced resolution---especially in the low energy range $0.5-2$~keV, see Section~\ref{sec:simulations}---this is the first time to our knowledge that such a detection is reported for the vicinity of Tycho's SNR shock wave in the X-rays.

Within the 'non-thermal+ejecta+ISM' class, the bayesian selection does not allow to unambiguously pick a specific model as best model.
In order to simulate the expected high-resolution X-ray spectrum of the shock regions of Tycho we use the  \texttt{Tbabs(powerlaw+nei+vnei)} model
as template, and we chose Region~2 as it is the region exacerbating the strongest synchrotron emission and the most featureless spectrum (it is therefore the region where increased spectral resolution is of most interest). The model and its components are displayed in Figure~\ref{fig:powpshockvnei}.

\section{Athena/X-IFU simulations}
\label{sec:simulations}

\subsection{Qualitative characterization of shock spectra}

We presented in previous sections our analysis on Tycho's SNR shock spectra. The Bayesian framework gave a new point of view on the data compared to previous studies and allowed us to select the most adequate model(s) at representing the spectra over all analysed spatial regions. Based on Bayes factor and other metrics, we found that 'non-thermal+ejecta+ISM' models were the most adequate to fit the spectra. This provided a first insight on the value of the electron temperature $T_e$, given by the continuum of these models. As said in the introduction of this paper, this value is interesting in relation to the proton temperature $T_p$ and to their ration \Trat \citep{Ghavamian2013,Vink2015}.

However, given the moderate spectral resolution of the \textit{Chandra}/ACIS-I detector and the molecular contamination of the optical blocking filter, it is not possible to make such diagnostics with this detector or similar ones (e.g., \textit{XMM-Newton}/pn). This is particularly relevant in this case given the low $n_et$ ($\sim10^{9}$~cm$^{-3}$s) values measured: emission lines are strongly shifted to low energies (typically towards the range 0.5 - 2~keV, see the \texttt{nei} components in Figure~\ref{fig:powpshockvnei} for examples). This partially explains the difficulty of detecting any contribution from the shocked ISM, especially with a frequentist approach as done in previous studies. The thermal emission lines would mainly reside at low energy, where both the low statistics and the spectral resolution do not allow to resolve them, while the high energy emission is dominated by the featureless non-thermal emission.

Future instruments such as \textit{Athena}/X-IFU will feature an increased efficient area and spectral resolution ($\leq$~3 eV in the 0.2 - 12 keV energy range). Such data could provide a clear view on this topic and allow us to confirm or deny this characterization of shock spectra. In the next subsections, we use our best Bayesian selected model, the \texttt{TBabs(powerlaw+nei+vnei)} model (see Section~\ref{subsec:best_bayesian_selected_model}), as template to simulate X-IFU spectra of the shock regions in order to measure effects of the doppler broadening on the emission lines.

\subsection{Simulation parameters}
\label{subsec:simulation_parameters}


\begin{figure}
\centering
\includegraphics[width=\linewidth]{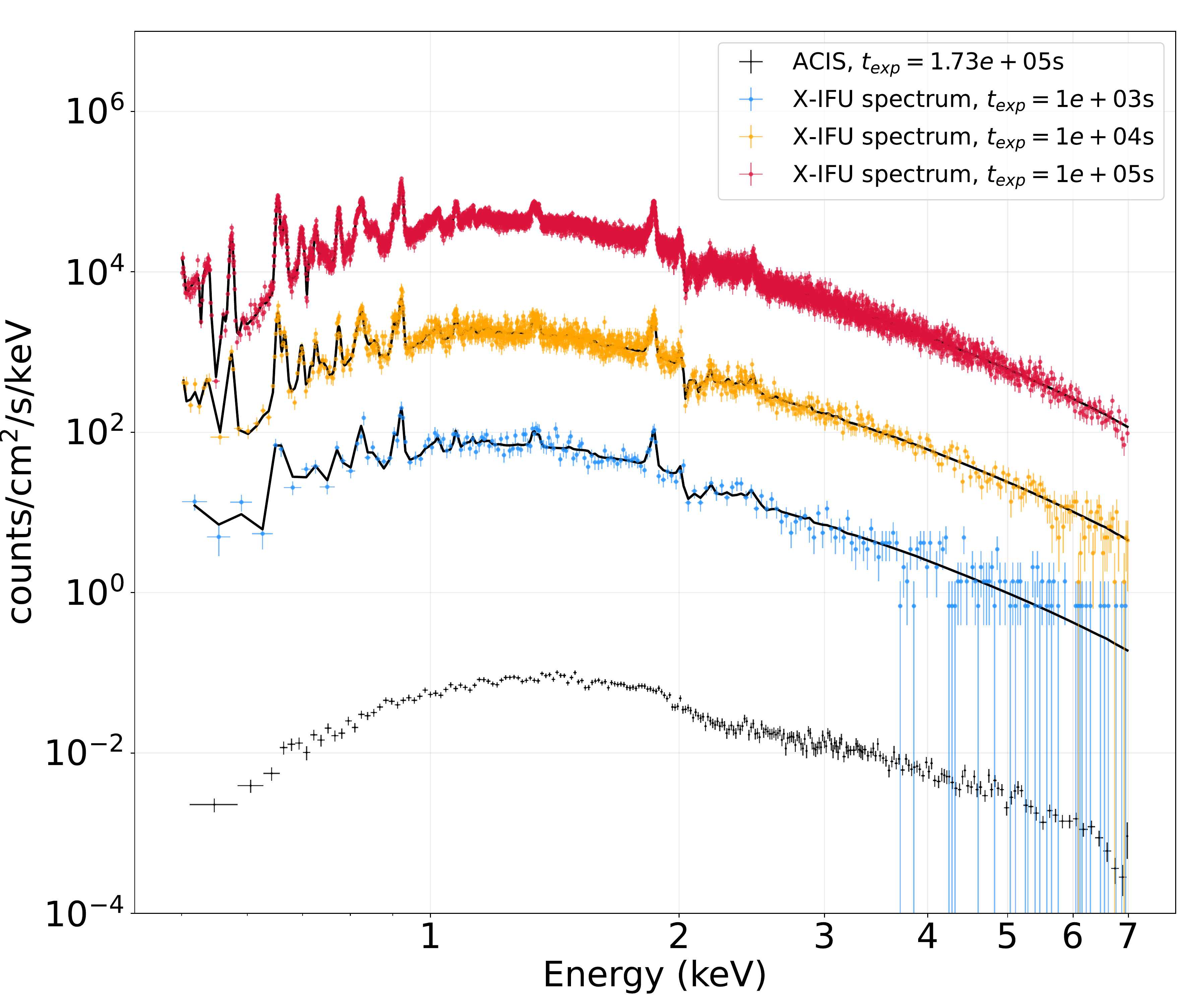}
\caption{Simulated X-IFU spectra with different exposure time values are displayed alongside the ACIS shock spectrum of Region~2 (black crosses). The normalization of the ACIS spectrum is the original, but the X-IFU spectra have been renormalized by a factor ten for display purposes. The black lines are fitted \texttt{TBabs(powerlaw+nei+vnei)} models with unfrozen $\sigma_E$ (see text of Section~\ref{subsec:method} for details).}
\label{fig:sim_xifu_spectra_grid}
\end{figure}

The Doppler broadening of emission lines is modeled by the \texttt{XSPEC} convolution model, \texttt{gsmooth}, which implements a simple Gaussian smoothing of the spectrum. All subsequent simulated spectra use the \texttt{TBabs(powerlaw+nei+vnei)} model, with posterior distributions computed with \texttt{BXA} in Region~2. The actual model used for the simulation is therefore given by \texttt{TBabs(gsmooth(powerlaw+nei+vnei))}. This adds two additional parameters, the Gaussian standard deviation $\sigma_E$ and the scaling index of $\sigma$ with energy, denoted $\alpha$. The last is usually frozen to unity. In the strong shock limit, $\sigma_E$ is given by
$$\sigma_E = \sqrt{\frac{3}{16}}\frac{E_0}{c}V_s,$$

\noindent
with $E_0=6$~keV here. To simulate spectra, we draw random values from the posterior distributions sampled in Section~\ref{sec:statistical_analysis}. In the case of the $\sigma_E$, we draw random shock velocity values from a Gaussian distribution with mean $V_s$ from \citet{Williams2016} (see Table~\ref{tab:williams_temperatures}), and with standard deviation the statistical error on the measurement from the same paper. This gives roughly $\sigma_E$ in the range $\sim28-30$~eV. We chose to draw 100 parameter sets.

We simulated spectra using the \texttt{XSPEC} command \textit{fakeit}. The Auxiliary Response File (ARF), Redistribution Matrix File (RMF) and background files are taken from the publicly available online version of the X-IFU simulation tools\footnote{http://x-ifu.irap.omp.eu/resources/for-the-community} \citep[see][for more details]{Barret2018}. In order to investigate the sensitivity of X-IFU in detecting the doppler broadening at different exposure times, we simulated three different spectra with exposure times of $10^3$, $10^4$ and $10^5$~s respectively, for each parameter set, leading to a total of 300 simulated spectra. Examples of such spectra are displayed in Figure~\ref{fig:sim_xifu_spectra_grid}.


\subsection{Method}
\label{subsec:method}

\begin{table*}
    \begin{tabular}{c c c c c c c c}
        \hline\hline\\
        Exposure time & $<C_{r, \mathrm{A}}>$ &  $<C_{r, \mathrm{B}}>$ &  $<C_{r, \mathrm{C}}>$ & $<\Delta C_{\mathrm{A-B}>}$ & $<\Delta C_{\mathrm{B-C}}>$ & $<$~Confidence interval~$>$ & $<e_r>$ \\
        $\left[\mathrm{ks}\right]$ & & & & & & $\left[\%\right]$ & $\left[\%\right]$\\\\
        \hline \\
        1 & \ama{0.44} & \ama{0.43} & \ama{0.43} & \ama{159} & \ama{0.81} & \ama{63.1} & \ama{$0.33^{+4.9}_{-19.0}$} \\
        10 & \ama{0.88} & \ama{0.78} & \ama{0.77} & \ama{1736} & \ama{0.84} & \ama{64.1} & \ama{$0.77^{+2.2}_{-2.4}$} \\
        100 & \ama{2.18} & \ama{1.04} & \ama{1.03} & \ama{15673} & \ama{0.94} & \ama{66.7} & \ama{$0.04^{+1.7}_{-2.3}$} \\\\

        \hline
        \\
    \end{tabular}
    \caption{Fit results averaged over 100 simulated X-IFU spectra for each exposure time bin. From left to right: exposure time in ks, averaged reduced \textit{C-stat} for model A, averaged reduced \textit{C-stat} for model B, averaged reduced \textit{C-stat} for model C, averaged $\Delta$\textit{C-stat} between models A and B, averaged $\Delta$\textit{C-stat} between models B and C and, the corresponding averaged confidence interval in percent, and the averaged relative error on the measure of $\sigma_E$ with the uncertainties given by the width of the error distribution. All averages are medians. See text for details.}
    \label{tab:xifu_cstat_tab}
\end{table*}

Contrary to analysis of the observed spectra, we are not interested in refined parameter space exploration but rather want to estimate the possibility to detect thermal Doppler broadening in the 300 simulated spectra, in order to infer ion temperatures. The complete Bayesian approach for each individual spectrum is therefore not necessary, and we go back to the faster frequentist approach. Fits are performed with the \texttt{XSPEC} Levenberg-Marquadt algorithm and with the \textit{C-stat} as minimization statistics.

We fit the simulated spectra with three different models: A) a \texttt{TBabs(powerlaw+nei+vnei)} model, i.e. without the \texttt{gsmooth} component, B) a \texttt{TBabs(gsmooth(powerlaw+nei+vnei))} model with $\sigma_E$ frozen to the true value drawn in Section~\ref{subsec:simulation_parameters}, C) same as model B but with unfrozen $\sigma_E$. In each case, we use the reduced \textit{C-stat} to estimate goodness-of-fit, denoted $C_r=\mathrm{\textit{C-stat}/K}$ with $K$ the number of degrees of freedom.

We computed for each spectrum the $\Delta$\textit{C-stat} (hereafter $\Delta C$ with $\Delta C_{\mathrm{X-Y}}=$~\textit{C-stat}$_{\mathrm{X}}-$\textit{C-stat}$_{\mathrm{Y}}$) between models A and B, as well as B and C. As $\Delta C$ is approximately $\Delta \chi^2$ distributed, $\Delta C_{\mathrm{A-B}}$ allow us to constrain how significant is the addition of the \texttt{gsmooth} component into the simulated spectra, and how much it impacts the fits. A low value for $\Delta C_{\mathrm{A-B}}$ would mean a low significance of the detection of Doppler broadening, as any model without broadening would give similar goodness-of-fit. In addition, $\Delta C_{\mathrm{B-C}}$ gives the confidence interval at which the true value of $\sigma_E$ is recovered. The last metric we used is the relative error $e_r$ in percent, which is defined as

$$
e_r = 100\times\frac{\sigma_{E,\mathrm{B}} - \sigma_{E,\mathrm{C}}}{\sigma_{E,\mathrm{B}}},
$$

\noindent
and measure the accuracy of the fits when it comes to measure the value of $\sigma_E$.

\subsection{Constraints on Doppler broadening}
\label{subsec:constraints_on_doppler_broadening}


\begin{figure}
\centering
\includegraphics[width=\linewidth]{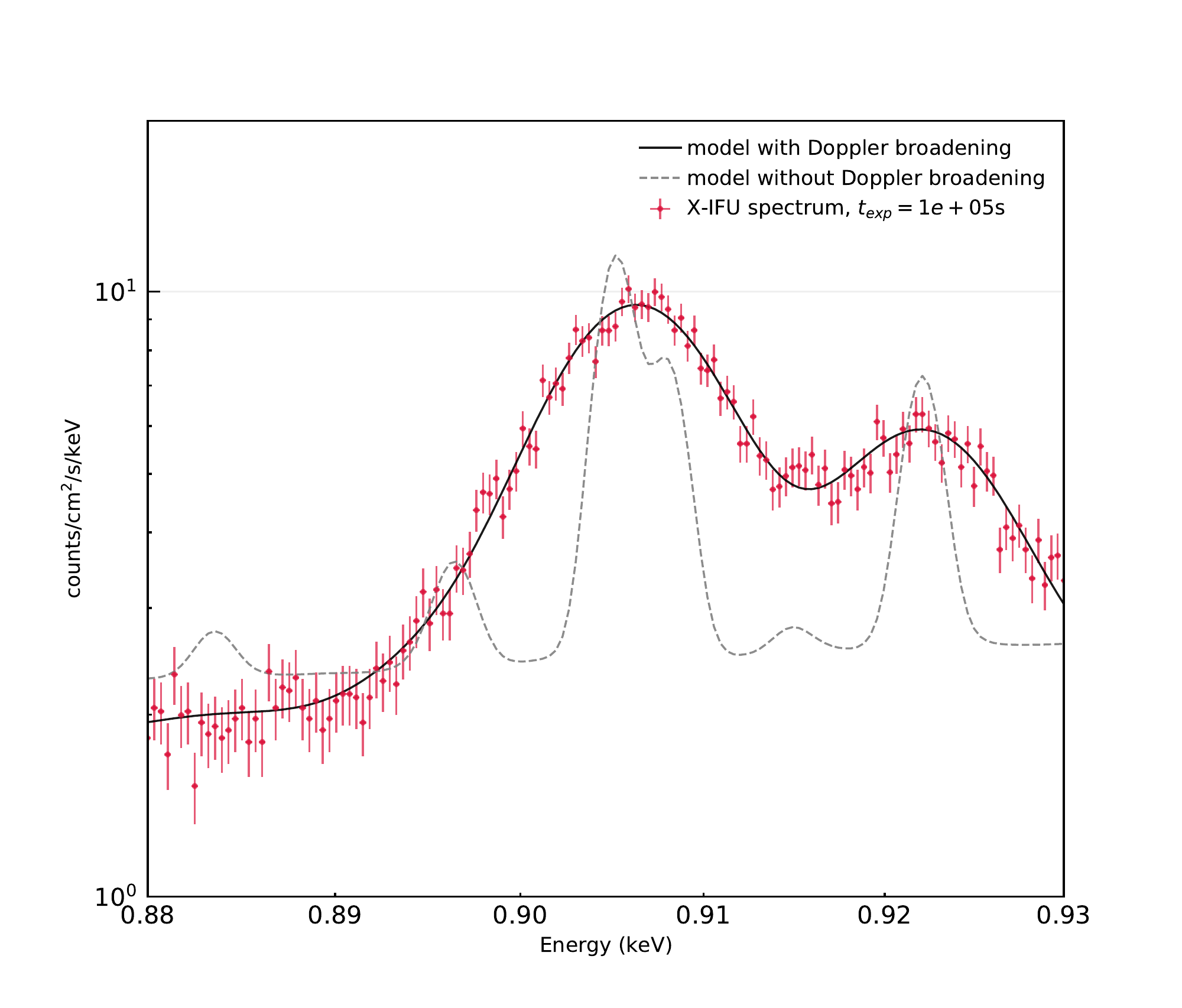}
\caption{Effect on the emission line of the inclusion of Doppler broadening in the fitted model. Red crosses are a simulated X-IFU spectrum with $\sigma_E=29$~eV and an exposure time of 100~ks, the black line is the fitted model with Doppler broadening, and the grey dashed line the model without.}
\label{fig:diff_el_gsmooth}
\end{figure}

For each of the three models and for each exposure time bin, we computed the median\footnote{One or two fits were not converging correctly, strongly skewing the distributions of $C_r$ and others metrics. Therefore we used the median and the median absolute variation (MAD) rather than the mean and standard deviation here, as it is more robust to outliers. } reduced \textit{C-stat} denoted $<C_r>$, as well as $<\Delta C_{\mathrm{A-B}}>$ and $<\Delta C_{\mathrm{B-C}}>$. We then computed confidence interval in percent corresponding to $<\Delta C_{\mathrm{B-C}}>$.  In the same way, we compute for each exposure time bin the median of the distribution $<e_r>$, as well as the positive and negative error bars given by the median absolute variation. All values are displayed in Table~\ref{tab:xifu_cstat_tab}.

The low $<C_r>$ values for exposure times of 1 and 10~ks are not surprising: error bars are very large and spectra are overfitted. Making any clear statement based on the analysis of spectra with such wide error bars is impossible: many different models could equally satisfyingly describe the same spectrum. This is not the case for spectra with exposure times of 100~ks. On the other hand, while $<C_{r, \mathrm{A}}>$ is quite similar to $<C_{r, \mathrm{B}}>$ and $<C_{r, \mathrm{C}}>$ for 1 and 10~ks spectra, it is much higher (\ama{$\sim2.18$}) for 100~ks spectra, indicating that models without the \texttt{gsmooth} component are much worse at fitting the spectra when given enough information. This is not the case of models B and C, which give almost perfect reduced \textit{C-stat} in average in this case.

This is further confirmed by looking at $<\Delta C_{\mathrm{A-B}}>$ which is very high. The improvement in goodness-of-fit when adding the \texttt{gsmooth} component to the models is excessively significant, and is due to the pristine spectral resolution of emission lines of X-IFU at low energies. This creates strong residuals when the model is not able to fit their width accurately, and results in very bad fits (see Figure~\ref{fig:diff_el_gsmooth}). In the opposite way, the $<\Delta C_{\mathrm{B-C}}>$ are very low, corresponding to confidence intervals of \ama{$\sim 65\%$}. This highlights excellent performance by model C, which shows that, in average, fitted $\sigma_E$ values approach very significantly the true value.

The relative error are also extremely low, showing how accurately the Doppler broadening is reproduced by model C. Please note however that while $<e_r>$ is very low (less than 1\% for each exposure time), the actual $e_r$ distributions are quite broad. Especially the fitted models have a tendency to overestimate the value of $\sigma_E$, with MAD higher (up to \ama{$\sim19\%$}) for negative relative errors. In the worst case, such errors would propagate relative errors of $\sim40\%$ on $T_i$\footnote{The ion temperature is given by $T_i = m_i \left( \sigma_E c / E_0 \right)^2$.}.

An interesting point to make here is the fact that while spectra with exposure times 1 and 10~ks are overfitted by model C, it still recovers the Doppler broadening, with relative error almost as good as 100~ks spectra. This suggests that observation times as short as 1~ks could be enough to settle the presence of emission lines at low energy and make measurements of Doppler broadening. Of course, extended observation times would still be needed to accurately model all its properties. In any case, these simulated spectra show how impactful X-IFU will be in unveiling the properties of thermal components in the vicinity of SNR blast waves, whether it is shocked ISM or ejecta.

\section{Conclusions}
\label{sec:conclusions}

In this work we report on an extended Bayesian analysis of the featureless spectra of five thin regions located at various points close to the Tycho's SNR shock wave, using $Chandra$ archive data. Our goal was to detect shocked ISM emission in the spectra, or at least to put constraints on its characteristics by performing a spatially resolved spectral analysis of previously analysed Chandra archival data with a more sophisticated statistical tool, based on bayesian inference.

The computations were performed with the \texttt{BXA} package, which allows to sample the posterior likelihood distribution of models over the data and compute for each model the marginalized likelihood also called Bayesian evidence (see Section~\ref{subsec:bayesian_xray_analysis}). The Bayesian evidences (i.e. \DL~values) were then used as the main metric to compare and select models in a simple way. This was the first time such an approach was applied to the study of thermal emission in the vicinity of Tycho's SNR blast wave.

We first compared output from \texttt{BXA} for single component models with results from literature, and found very good matches (see Section~\ref{sec:single_component_models}). We then applied it to more complicated cases, with composite models featuring both a thermal component and a non-thermal one (see Section~\ref{sec:bicomponent_models}). In the case of non-thermal+ISM models, we showed that the parameter spaces of such a class of models display complex shapes, with multiple likelihood maxima. We also showed the main modes of some of these models to be not physically possible, similarly to previous findings in the literature \citep{Cassam2007}. By adding constraints to the model parameters, we explored regions of these model parameter spaces less statistically plausible, but physically acceptable. However a better description of the spectra were given by non-thermal+ejecta models with free abundances, which were able to better reproduce emission lines in the spectra, even though some faint spectral line features were still not correctly reproduced in the low-energy band. We then added a third component to our models, and sampled the posteriors of non-thermal+ejecta+ISM models. We 
find that this class of models is
significantly better at describing the data than single-component and two-components models according to Bayes factors and other metrics, providing the first proof for shocked ISM emission in the shock front of Tycho's SNR.


We detected shocked ISM in all five regions, which allowed us to study spatial variations of its properties. We found rather low electron temperatures ($kT_{\mathrm{e,ISM}}=0.96^{+1.33}_{-0.51}$~keV) through all regions (see Figure~\ref{fig:plot_tricomponent_values_3by2}). These temperatures are lower than the mean shock temperature ($\sim12$~keV) by an order of magnitude. If the ion temperature is of the same order of magnitude than the mean shock velocity, this indicates a temperature ratio of the order of 0.1, which is far from the asymptotic $kT_{\rm e}/kT_{\rm p}=m_{\rm e}/m_{\rm p}$ in the strong shock limit and could indicate moderate energy transfer between ions and electrons. However, these are only assumptions, and ion temperature measurements are needed to settle this question.

From the ionization timescales of the shocked ISM and velocity measurements from \citet{Williams2016}, we were able to derive ambient electron densities (Figure~\ref{fig:plot_ne}). We find an average (excluding Region~3 in which the upper limit of $n_\mathrm{e}t$ is not constrained) ambient electron density $n_{\rm e}=0.32^{+0.23}_{-0.15}$~cm$^{-3}$, in agreement with the upper limit of $0.7$~cm$^{-3}$ reported by \citet{Cassam2007}. \ama{We find no significant difference in ambient density between the north regions ($n_{\mathrm{e}}=0.33^{+0.19}_{-0.15}$~cm$^{-3}$) and the southwest regions  $n_{\mathrm{e}}=0.31^{+0.29}_{-0.13}$~cm$^{-3}$), and we are not able to confirm the presence of the density gradient around Tycho'SNR found in infrared measurements by \citet{Williams2013}.}

In order to test the robustness of our analysis, we added several different synchrotron cutoff models in the region with the strongest non-thermal emission (e.g. Region~2, see Section~\ref{subsec:extended_analysis_of_region_2} for details). Not only this allowed us to compare more models, but also to evaluate how the thermal component properties interact with the add-on of cutoffs. We found that such models were not giving significantly better descriptions in terms of Bayesian evidence than composite models with a regular power-law. Following Occam's rule, the cutoff is therefore not necessary to describe the spectra. This shows that the presence of a faint thermal component in the spectra appears anti-correlated with the presence of a cutoff for the non-thermal emission.  

We showed that the low ionization timescale and electron temperature are shifting shocked ISM emission lines to low energy range, where \textit{Chandra}/ACIS sensitivity drops. In Section~\ref{sec:simulations} we showed that the future instrument X-IFU onboard \textit{Athena} will be able to resolve such low energy lines. Therefore, we simulated X-IFU spectra using our best Bayesian model as template, including also a \texttt{gsmooth} component to model thermal broadening. These simulations allowed us to estimate the possibility for thermal Doppler broadening measurement with X-IFU. We were able to measure Doppler broadening with average relative errors lower than $1\%$ and recovering the actual broadening with confidence intervals around $50\%$.

Overall, these results provide very promising  perspectives for future X-IFU observations of SNRs, not only to definitely settle the question on the presence of a thermal component in the vicinity of SNR shocks, but also to measure quantities such as the ion temperature which, in conjunction with the electron temperature, are essential to our understanding of collisionless shocks. 

\begin{acknowledgments}
Acknowledgments. We would like to thank the anonymous referee for the quality and the depth of the comments made about this work, which allowed us to refine greatly our analysis. AE and this work are supported by the research program Athena with project number 184.034.002, which is (partially) financed by the Dutch Research Council (NWO). JV and EG are supported by funding from the European Unions Horizon 2020 research and innovation programme under grant agreement No 101004131 (SHARP). This work has made use of the CANDIDE Cluster at the Institut d'Astrophysique de Paris and made possible by grants from the PNCG, CNES, and the DIM-ACAV.
\end{acknowledgments}

\bibliography{sample631}{}
\bibliographystyle{aasjournal}



\appendix
\restartappendixnumbering

\section{Additional metrics for model selection}

We display in this appendix the additional model comparison metrics for Sections~\ref{sec:single_component_models}, \ref{sec:bicomponent_models} and \ref{sec:three_components} (see main text for details).

\begin{table*}[!h]
\caption{\textbf{Akaike information criterion for every models of the five regions.}}
\label{tab:aic}
\centering
\resizebox{\textwidth}{!}{
\begin{tabular}{l l c c c c c c}
\hline\hline\\\\
& & \multicolumn{6}{c}{AIC$_{\mathrm{min}}$-AIC} \\\\\cline{3 - 8}\\
& Model & Region 1 & Region 2 & Region 3 & Region 4 & Region 5 & Average \\\\
\hline\\
\multirow{4}{*}{Single non-thermal} 
& \texttt{TBabs(powerlaw)} & 631.81  & 159.24  & 422.64  & 538.11  & 501.11  & 450.58 \\ 
& \texttt{TBabs(srcut)} & 563.02  & 121.74  & 361.25  & 492.76  & 459.79  & 399.71 \\ 
& \texttt{TBabs(cutoffpl)} & 471.86  & 110.11  & 306.86  & 433.55  & 415.25  & 347.53 \\ 
& \texttt{TBabs(sqrtcutoffpl)} & 472.02  & 99.89  & 296.56  & 430.76  & 407.91  & 341.43 \\ 
\\
\hline
\\
\multirow{3}{*}{Single ISM} 
& \texttt{TBabs(nei)} & 533.06  & 303.28  & 525.42  & 545.51  & 445.94  & 470.64 \\ 
& \texttt{TBabs(pshock)} & 527.13  & 274.35  & 480.47  & 529.13  & 435.67  & 449.35 \\ 
& \texttt{TBabs(npshock)} & 527.15  & 274.36  & 480.50  & 529.14  & 435.67  & 449.36 \\ 
\\
\hline
\\
\multirow{1}{*}{Single Ejecta} 
& \texttt{TBabs(vnei)} & 107.66  & 110.99  & 176.25  & 115.41  & 91.51  & 120.36 \\ 
\\
\hline
\\
\multirow{1}{*}{Non-thermal + thermal} 
& \texttt{TBabs(powerlaw+nei)} & 397.73  & 85.08  & 294.62  & 382.94  & 326.91  & 297.46 \\ 
\\
\hline
\\
\multirow{1}{*}{Non-thermal + ejecta} 
& \texttt{TBabs(powerlaw+vnei)} & 29.35  & 30.45  & 15.52  & 22.52  & 22.35  & 24.04 \\ 
\\
\hline
\\
\multirow{1}{*}{Non-thermal + ejecta + ISM} 
& \texttt{TBabs(powerlaw+nei+vnei)} & 0.00  & 0.00  & 0.00  & 0.00  & 0.00  & 0.00 \\ 
\\
\hline
\end{tabular}}
\tablenotetext{}{Notes. For each region (column) the AIC of every model is normalized to the best one in that region (column). In the last right column is shown the AIC for each model averaged over all regions, normalized by the best average AIC. The best model for each region has a value of zero.}
\end{table*}

\begin{table*}
\caption{\textbf{Bayesian information criterion for every models of the five regions.}}
\label{tab:bic}
\centering
\resizebox{\textwidth}{!}{
\begin{tabular}{l l c c c c c c}
\hline\hline\\\\
& & \multicolumn{6}{c}{BIC$_{\mathrm{min}}$-BIC} \\\\\cline{3 - 8}\\
& Model & Region 1 & Region 2 & Region 3 & Region 4 & Region 5 & Average \\\\
\hline\\
\multirow{4}{*}{Single non-thermal} 
& \texttt{TBabs(powerlaw)} & 691.51  & 217.85  & 496.17  & 604.65  & 567.82  & 515.60 \\ 
& \texttt{TBabs(srcut)} & 622.72  & 180.35  & 434.79  & 559.30  & 526.49  & 464.73 \\ 
& \texttt{TBabs(cutoffpl)} & 523.46  & 160.62  & 372.29  & 491.99  & 473.86  & 404.45 \\ 
& \texttt{TBabs(sqrtcutoffpl)} & 523.63  & 150.40  & 362.00  & 489.20  & 466.52  & 398.35 \\ 
\\
\hline
\\
\multirow{3}{*}{Single ISM} 
& \texttt{TBabs(nei)} & 584.67  & 353.79  & 590.86  & 603.95  & 504.55  & 527.56 \\ 
& \texttt{TBabs(pshock)} & 578.74  & 324.86  & 545.91  & 587.57  & 494.28  & 506.27 \\ 
& \texttt{TBabs(npshock)} & 578.76  & 324.87  & 545.94  & 587.58  & 494.28  & 506.28 \\ 
\\
\hline
\\
\multirow{1}{*}{Single Ejecta} 
& \texttt{TBabs(vnei)} & 94.50  & 96.74  & 176.92  & 109.08  & 85.35  & 112.52 \\ 
\\
\hline
\\
\multirow{1}{*}{Non-thermal + thermal} 
& \texttt{TBabs(powerlaw+nei)} & 433.14  & 119.40  & 343.86  & 425.19  & 369.33  & 338.19 \\ 
\\
\hline
\\
\multirow{1}{*}{Non-thermal + ejecta} 
& \texttt{TBabs(powerlaw+vnei)} & 0.00  & 0.00  & 0.00  & 0.00  & 0.00  & 0.00 \\ 
\\
\hline
\\
\multirow{1}{*}{Non-thermal + ejecta + ISM} 
& \texttt{TBabs(powerlaw+nei+vnei)} & 11.13  & 10.03  & 24.96  & 17.96  & 18.13  & 16.44 \\ 
\\
\hline
\end{tabular}}
\tablenotetext{}{Notes. For each region (column) the BIC of every model is normalized to the best one in that region (column). In the last right column is shown the BIC for each model averaged over all regions, normalized by the best average BIC. The best model for each region has a value of zero. Please note that according to BIC, the 'non-thermal+ejecta' model performs better than the three component model 'non-thermal+ejecta+ISM' because it is biased by the way the abundances were handled in our analysis (see Section~\ref{subsec:comparison_model_tri} for details).}
\end{table*}

\begin{table*}
\caption{\textbf{Difference in \textit{C-stat} for every models of the five regions.}}
\label{tab:deltac}
\centering
\resizebox{\textwidth}{!}{
\begin{tabular}{l l c c c c c c}
\hline\hline\\\\
& & \multicolumn{6}{c}{$\Delta$\textit{C-stat} } \\\\\cline{3 - 8}\\
& Model & Region 1 & Region 2 & Region 3 & Region 4 & Region 5 & Average \\\\
\hline\\
\multirow{4}{*}{Single non-thermal} 
& \texttt{TBabs(powerlaw)} & 643.81  & 171.24  & 434.64  & 550.11  & 513.11  & 462.58 \\ 
& \texttt{TBabs(srcut)} & 575.02  & 133.74  & 373.25  & 504.76  & 471.79  & 411.71 \\ 
& \texttt{TBabs(cutoffpl)} & 481.86  & 120.11  & 316.86  & 443.55  & 425.25  & 357.53 \\ 
& \texttt{TBabs(sqrtcutoffpl)} & 482.02  & 109.89  & 306.56  & 440.76  & 417.91  & 351.43 \\ 
\\
\hline
\\
\multirow{3}{*}{Single ISM} 
& \texttt{TBabs(nei)} & 543.06  & 313.28  & 535.42  & 555.51  & 455.94  & 480.64 \\ 
& \texttt{TBabs(pshock)} & 537.13  & 284.35  & 490.47  & 539.13  & 445.67  & 459.35 \\ 
& \texttt{TBabs(npshock)} & 537.15  & 284.36  & 490.50  & 539.14  & 445.67  & 459.36 \\ 
\\
\hline
\\
\multirow{1}{*}{Single Ejecta} 
& \texttt{TBabs(vnei)} & 101.66  & 104.99  & 170.25  & 109.41  & 85.51  & 114.36 \\ 
\\
\hline
\\
\multirow{1}{*}{Non-thermal + thermal} 
& \texttt{TBabs(powerlaw+nei)} & 403.73  & 91.08  & 300.62  & 388.94  & 332.91  & 303.46 \\ 
\\
\hline
\\
\multirow{1}{*}{Non-thermal + ejecta} 
& \texttt{TBabs(powerlaw+vnei)} & 19.35  & 20.45  & 5.52  & 12.52  & 12.35  & 14.04 \\ 
\\
\hline
\\
\multirow{1}{*}{Non-thermal + ejecta + ISM} 
& \texttt{TBabs(powerlaw+nei+vnei)} & 0.00  & 0.00  & 0.00  & 0.00  & 0.00  & 0.00 \\ 
\\
\hline
\end{tabular}}

\tablenotetext{}{Notes. For each region (column) the $\Delta$\textit{C-stat} of every model is normalized to the best one in that region (column). In the last right column is shown the $\Delta$\textit{C-stat} for each model averaged over all regions, normalized by the best average $\Delta$\textit{C-stat}. The best model for each region has a value of zero.}

\end{table*}

\section{Posterior distribution of three-component models for all regions}
\label{app:posterior_all_regions}
We display here the posterior distributions of the best bayesian selected three-component models for all regions (see Table~\ref{tab:regions_logz}).

\section{Bayesian evidence for all models in Region~2}

We display in this appendix additional content for Section~\ref{subsec:extended_analysis_of_region_2} (see main text for details).

\begin{figure*}
    \centering
    \includegraphics[width=\textwidth]{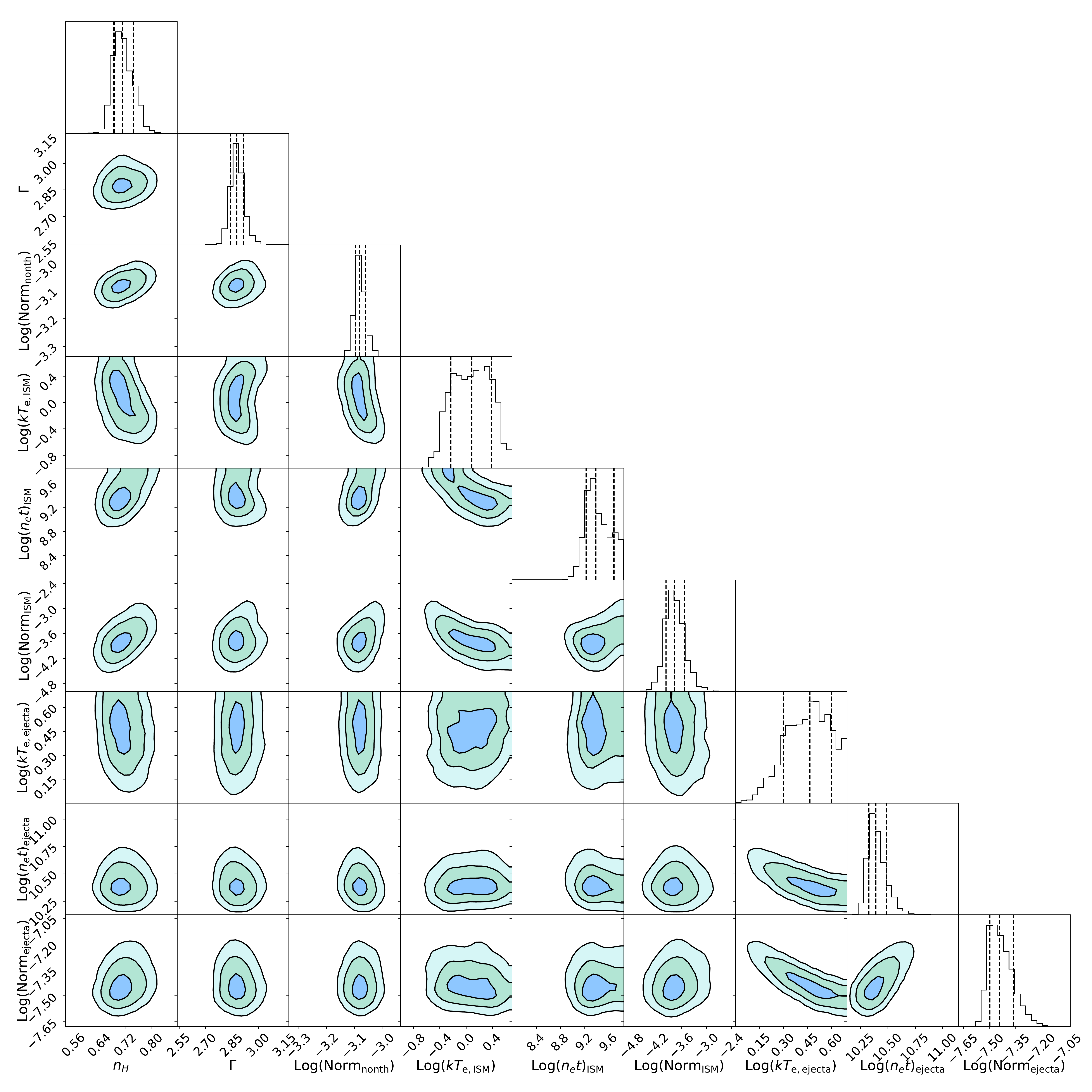}
    \caption{Posterior likelihood distribution of the \texttt{TBabs(powerlaw+nei+vnei)}~model on Region~1. The distributions have been sampled with the \texttt{BXA} package. Log-uniform priors have been set for all parameters, with the exception of the power-law photon index $\Gamma$ and the Hydrogen absorption $N_H$, which have received linear priors. The temperatures are given in keV, the normalization of the non-thermal component in keV$^{-1}$ cm$^2$ s$^{-1}$, the ionization timescales in cm$^{-3}$s and the normalizations of the thermal components in cm$^{-3}$. The top plot of each column shows the individual parameter histograms\ama{, with the 0.16, 0.5 and 0.84 percentiles showed as dashed vertical black lines.} The contours correspond (from darker to lighter blue) to $1\sigma$, $2\sigma$ and $3\sigma$ significance levels.}
    \label{fig:posterior_powneivnei_1}
\end{figure*}


\begin{figure*}
    \centering
    \includegraphics[width=\textwidth]{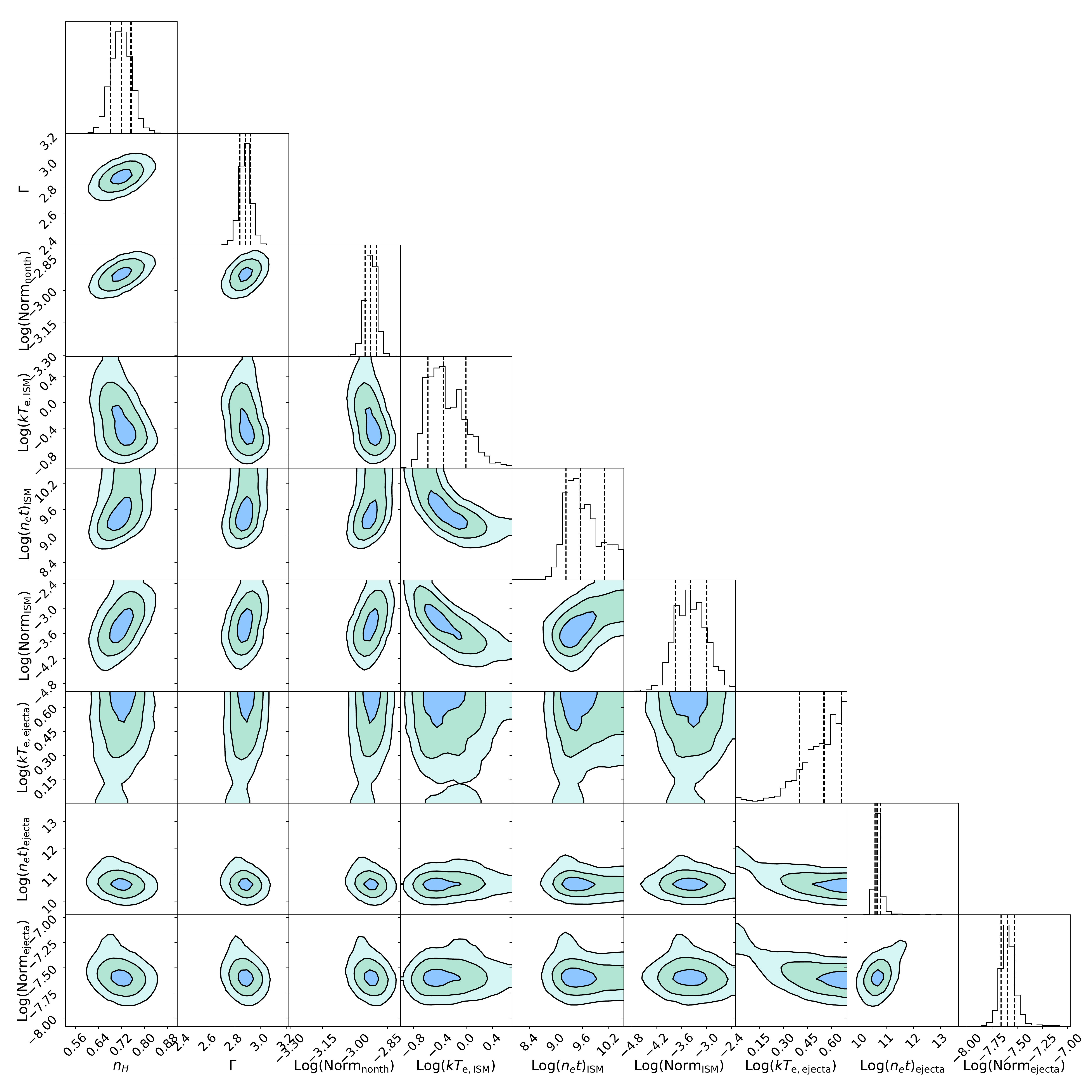}
    \caption{Same as for Figure~\ref{fig:posterior_powneivnei_1} but for Region~2.}
    \label{fig:posterior_powneivnei_2}
\end{figure*}


\begin{figure*}
    \centering
    \includegraphics[width=\textwidth]{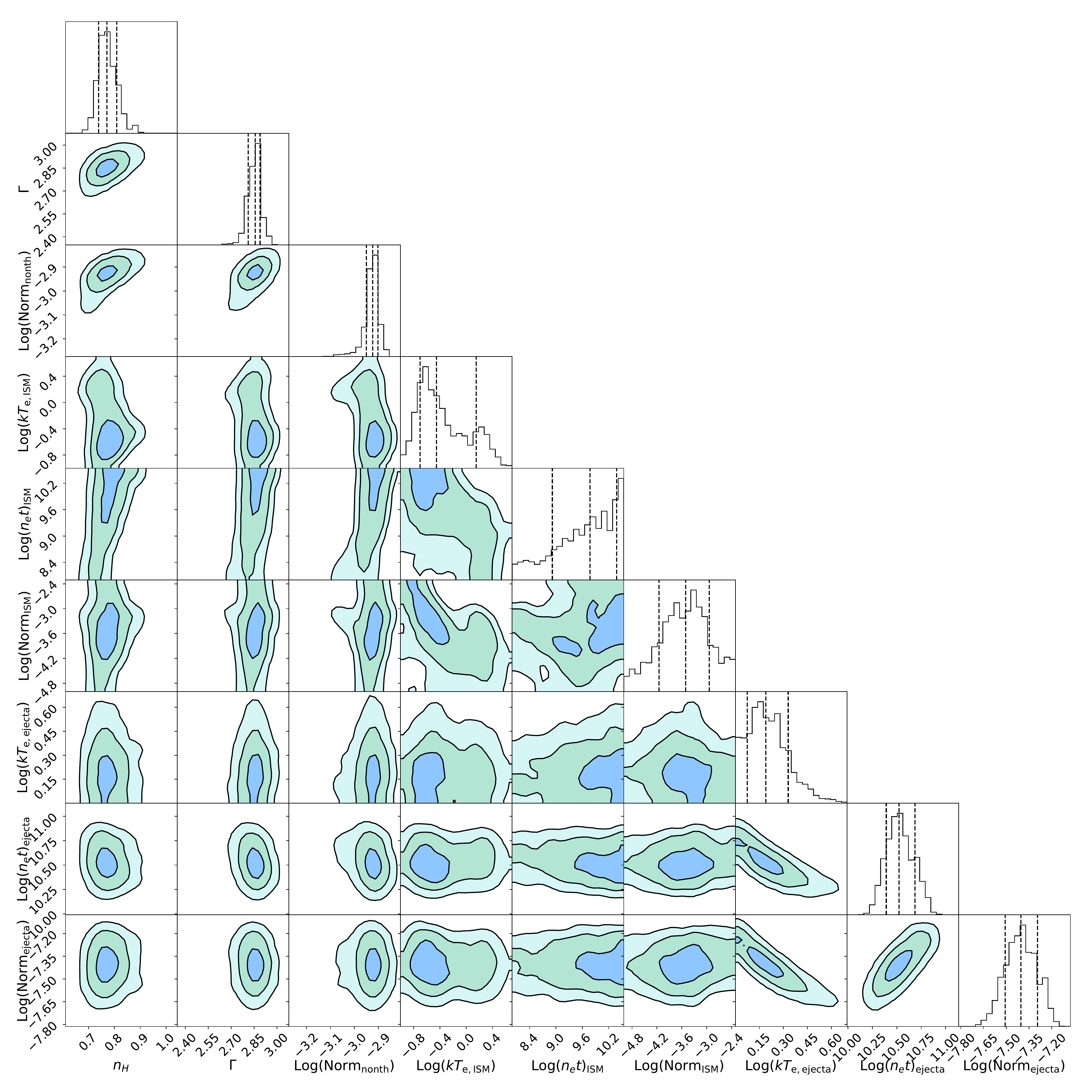}
    \caption{Same as for Figure~\ref{fig:posterior_powneivnei_1} but for Region~3.}
    \label{fig:posterior_powneivnei_3}
\end{figure*}


\begin{figure*}
    \centering
    \includegraphics[width=\textwidth]{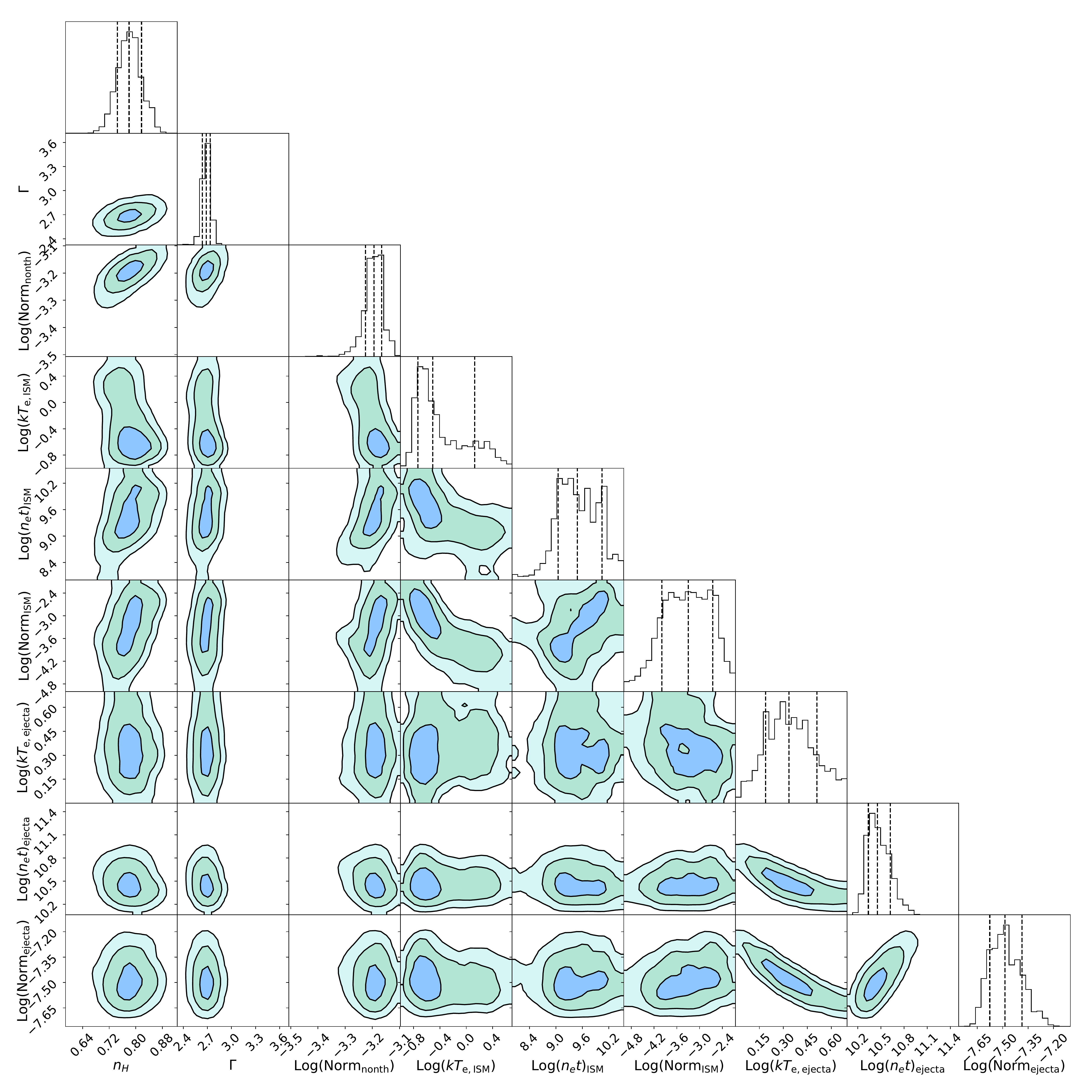}
    \caption{Same as for Figure~\ref{fig:posterior_powneivnei_1} but for Region~4.}
    \label{fig:posterior_powneivnei_4}
\end{figure*}


\begin{figure*}
    \centering
    \includegraphics[width=\textwidth]{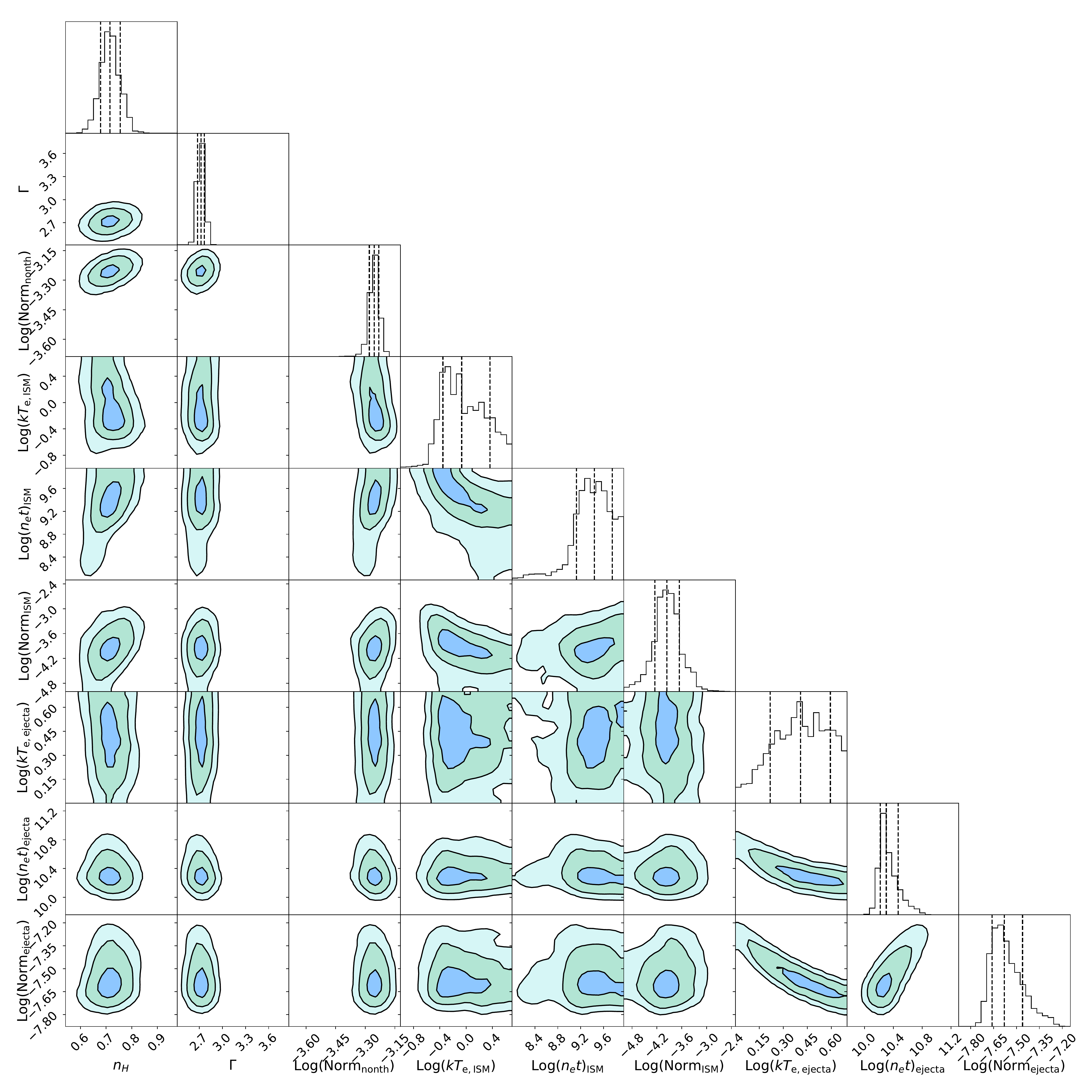}
    \caption{Same as for Figure~\ref{fig:posterior_powneivnei_1} but for Region~5.}
    \label{fig:posterior_powneivnei_5}
\end{figure*}


\begin{table*}
\caption{Bayes factors and other goodness-of-fit metrics for every model (including models with an energy cutoff) for Region~2.}
\label{tab:logz_table_reg2_all}
\centering
\resizebox{\textwidth}{!}{
\begin{tabular}{l l c c c c}
\hline\hline\\
& Model & $\log(z) - \log(z)_\mathrm{max}$ & $\Delta C$ & $AIC - AIC_{\mathrm{min}}$ & $BIC - BIC_{\mathrm{min}}$ \\\\
\hline\\ 
& \texttt{TBabs(powerlaw)} & -74.37  & 148.73  & 138.73  & 196.83 \\ 
& \texttt{TBabs(srcut)} & -57.08  & 114.16  & 104.16  & 162.26 \\ 
& \texttt{TBabs(cutoffpl)} & -51.96  & 103.92  & 95.92  & 145.91 \\ 
& \texttt{TBabs(sqrtcutoffpl)} & -45.14  & 90.28  & 82.28  & 132.28 \\ 
\\
\hline
\\
\multirow{3}{*}{Single thermal} 
& \texttt{TBabs(nei)} & -151.09  & 302.19  & 294.19  & 344.18 \\ 
& \texttt{TBabs(pshock)} & -135.21  & 270.42  & 262.42  & 312.41 \\ 
& \texttt{TBabs(npshock)} & -134.90  & 269.81  & 261.81  & 311.80 \\ 
\\
\hline
\\
\multirow{12}{*}{Non-thermal + ISM} 
& \texttt{TBabs(powerlaw+nei)} & -40.31  & 80.63  & 76.63  & 110.43 \\ 
& \texttt{TBabs(powerlaw+pshock)} & -30.87  & 61.73  & 59.73  & 85.44 \\ 
& \texttt{TBabs(powerlaw+npshock)} & -40.52  & 81.04  & 79.04  & 104.75 \\ 
& \texttt{TBabs(srcut+nei)} & -35.07  & 70.14  & 70.14  & 87.76 \\ 
& \texttt{TBabs(srcut+pshock)} & -35.24  & 70.48  & 70.48  & 88.09 \\ 
& \texttt{TBabs(srcut+npshock)} & -35.47  & 70.93  & 70.93  & 88.55 \\ 
& \texttt{TBabs(cutoffpl+nei)} & -36.30  & 72.60  & 72.60  & 90.22 \\ 
& \texttt{TBabs(cutoffpl+pshock)} & -36.34  & 72.67  & 72.67  & 90.29 \\ 
& \texttt{TBabs(cutoffpl+npshock)} & -36.52  & 73.05  & 73.05  & 90.66 \\ 
& \texttt{TBabs(sqrtcutoffpl+nei)} & -33.77  & 67.55  & 67.55  & 85.16 \\ 
& \texttt{TBabs(sqrtcutoffpl+pshock)} & -33.68  & 67.36  & 67.36  & 84.98 \\ 
& \texttt{TBabs(sqrtcutoffpl+npshock)} & -33.88  & 67.75  & 67.75  & 85.37 \\ 
\\
\hline
\\
\multirow{4}{*}{Non-thermal + ejecta} 
& \texttt{TBabs(powerlaw+vnei)} & -9.48  & 18.96  & 30.96  & 0.00 \\ 
& \texttt{TBabs(srcut+vnei)} & -15.70  & 31.41  & 25.41  & 67.31 \\ 
& \texttt{TBabs(cutoffpl+vnei)} & -16.25  & 32.50  & 28.50  & 62.30 \\ 
& \texttt{TBabs(sqrtcutoffpl+vnei)} & -14.08  & 28.15  & 24.15  & 57.96 \\ 
\\
\hline
\\
\multirow{12}{*}{Non-thermal + ejecta + ISM} 
& \texttt{TBabs(powerlaw+nei+vnei)} & -0.00  & 0.00  & 2.00  & 11.52 \\ 
& \texttt{TBabs(powerlaw+pshock+vnei)} & -0.86  & 1.72  & 1.72  & 19.33 \\ 
& \texttt{TBabs(powerlaw+npshock+vnei)} & 0.00  & 0.00  & 0.00  & 17.61 \\ 
& \texttt{TBabs(srcut+nei+vnei)} & -1.99  & 3.98  & 3.98  & 21.59 \\ 
& \texttt{TBabs(srcut+pshock+vnei)} & -7.08  & 14.17  & 14.17  & 31.78 \\ 
& \texttt{TBabs(srcut+npshock+vnei)} & -5.64  & 11.29  & 11.29  & 28.90 \\ 
& \texttt{TBabs(cutoffpl+nei+vnei)} & -2.88  & 5.76  & 7.76  & 17.27 \\ 
& \texttt{TBabs(cutoffpl+pshock+vnei)} & -3.23  & 6.47  & 8.47  & 17.99 \\ 
& \texttt{TBabs(cutoffpl+npshock+vnei)} & -7.27  & 14.54  & 16.54  & 26.06 \\ 
& \texttt{TBabs(sqrtcutoffpl+nei+vnei)} & -0.22  & 0.43  & 2.43  & 11.95 \\ 
& \texttt{TBabs(sqrtcutoffpl+pshock+vnei)} & -0.70  & 1.39  & 3.39  & 12.91 \\ 
& \texttt{TBabs(sqrtcutoffpl+npshock+vnei)} & -4.93  & 9.86  & 11.86  & 21.38 \\ 
\\
\hline
\end{tabular}}
\end{table*}

\end{document}